\documentclass[longauth]{aa}

\usepackage{graphicx}
\usepackage{txfonts}
\usepackage{hyperref}
\hypersetup{
    colorlinks=true,
    linkcolor=blue,
    urlcolor=cyan
    }

\begin{document}

   \title{DESHIMA~2.0: A 200-400 GHz Ultra-wideband Integrated Superconducting Spectrometer}

%   \subtitle{ultra-wideband integrated superconducting spectrometer}

   \author{
   		%K. Karatsu \inst{1}\fnmsep\thanks{Corresponding author email: k.karatsu@sron.nl},
		K.~Karatsu \thanks{Corresponding author email: k.karatsu@sron.nl} \inst{1},
   		A.~Endo \inst{2},
		A.~Moerman \inst{2},
		S.~J.~C.~Yates \inst{1},
		R.~Huiting \inst{1},
		A.~Pascual~Laguna \inst{3},
		S.~Dabironezare \inst{1,2},
		V.~Murugesan \inst{1},
		D.~J.~Thoen \inst{1},
		%H.~Akamatsu \inst{1},
		%T.~J.~L.~C.~Bakx \inst{4,7,15},
		B.~T.~Buijtendorp \inst{2},
		S.~Cray \inst{4},
		K.~Fujita \inst{5},
		S.~H\"ahnle \inst{2},
		S.~Hanany \inst{4},
		R.~Kawabe \inst{6},
		%T.~Kitayama \inst{8},
		K.~Kohno \inst{7,8},
		%K.~Konishi \inst{9},
		L.~H.~Marting \inst{2},
		T.~Matsumura \inst{9},
		S.~Nakatsubo \inst{10},
		L.~G.~G.~Olde Scholtenhuis \inst{2},
		T.~Oshima \inst{6},
		M.~Rybak \inst{2,11},
		%H.~Sakurai \inst{9},
		F.~Steenvoorde \inst{2},
		R.~Takaku \inst{9},
		T.~Takekoshi \inst{12},
		Y.~Tamura \inst{13},
		A.~Taniguchi \inst{12,13},
		P.~P.~van~der~Werf \inst{11},
          \and
          	J.~J.~A.~Baselmans \inst{1,2,14}
          }

   \institute{
   		%1
   		SRON - Space Research Organisation Netherlands,
   		Niels Bohrweg 4, 2333 CA, Leiden, The Netherlands
	\and
		%2
		Faculty of Electrical Engineering, Mathematics and Computer Science, Delft University of Technology,
		Mekelweg 4, 2628 CD, Delft, The Netherlands
         \and
         	%3
         	Centro de Astrobiolog\'{i}a (CAB), INTA-CSIC,
		Ctra. de Torrej\'{o}n a Ajalvir km 4, 28850, Torrej\'{o}n de Ardoz, Spain
%	\and
%		%4
%		Department of Space, Earth, \& Environment, Chalmers University of Technology,
%		Chalmersplatsen 4 412 96 Gothenburg, Sweden
	\and
		%4
		School of Physics and Astronomy, University of Minnesota/Twin Cities,
		116 Church Street, SE, Minneapolis, MN 55455, USA
	\and
		%5
		Department of Cosmosciences, Graduate School of Science, Hokkaido University,
		Kita-10 Nishi-8, Kita-ku, Sapporo, 060-0810, Japan
	\and
		%6
		National Astronomical Observatory of Japan (NAOJ),
		2-21-1, Osawa, Mitaka, Tokyo, 181-8588, Japan
%	\and
%		%8
%		Department of Physics, Toho University,
%		2-2-1 Miyama, Funabashi, Chiba, 274-8510, Japan
	\and
		%7
		Institute of Astronomy, Graduate School of Science, The University of Tokyo,
		2-21-1 Osawa, Mitaka, Tokyo, 181-0015, Japan
	\and
		%8
		Research Center for the Early Universe, Graduate School of Science, The University of Tokyo,
		7-3-1 Hongo, Bunkyo-ku, Tokyo, 113-0033, Japan
	\and
		%9
		Kavli Institute for the Physics and Mathematics of the Universe (Kavli IPMU), The University of Tokyo,
		5-1-5 Kashiwanoha, Kashiwa, Chiba, 277-8583, Japan
	\and
		%10
		Institute of Space and Astronautical Science, Japan Aerospace Exploration Agency (ISAS/JAXA),
		3-1-1 Yoshinodai, Chuo-ku, Sagamihara, Kanagawa, 252-5210, Japan
	\and
		%11
		Leiden Observatory, Leiden University,
		PO Box 9513, 2300 RA, Leiden, The Netherlands
	\and
		%12
		Kitami Institute of Technology,
		165 Koen-cho, Kitami, Hokkaido, 090-8507, Japan
	\and
		%13
		Department of Physics, Graduate School of Science, Nagoya University,
		Furo-cho, Chikusa-ku, Nagoya, Aichi, 464-8602, Japan
	\and
		%14
		Physikalisches Institut, Universit\"{a}t zu K\"{o}ln,
		Z\"{u}lpicher Stra\ss e 77, 50937 Cologne, Germany
	}

   \date{Received xxxx, 2026; accepted xxxx, 2026}

% \abstract{}{}{}{}{}
% 5 {} token are mandatory
 
  \abstract
   {
   DESHIMA (Deep Spectroscopic HIgh-redshift MApper) is a broadband integrated superconducting spectrometer (ISS) for millimeter (mm) / sub-millimeter (sub-mm) wave astronomy based on Kinetic Inductance Detectors (KIDs).
   The first generation, DESHIMA 1.0, was successfully tested in 2017.
   This paper describes the upgrade to DESHIMA 2.0 and presents its characterization in laboratory settings.
   }
   {
   We aimed to enhance performance by increasing band coverage and efficiency while developing a fast sky-position chopper for efficient removal of atmospheric fluctuations.
   }
   {
   The instrument features NbTiN superconducting microstrip (MS) filters with low-loss a-SiC:H dielectric and an ultra-wideband leaky-wave antenna.
   A laboratory setup was designed, incorporating the cryostat housing cryogenic optics and ISS chip comprising 339 KIDs connected to MS filters tuned for (sub-)mm wave frequencies.
   Room-temperature mirrors on a hexapod stage allowed precise positioning and alignment of optical elements.
   The sky-position chopper was positioned on a motor-controlled stage for fine-tuned control over its position and alignment.
   Thanks to the multiplexing capability of KIDs, we could simultaneously measure multiple performance metrics across the entire frequency range.
   }
   {
   We showed that DESHIMA 2.0 achieved significant improvements in performance compared to its predecessor: measured instantaneous frequency coverage was $200 - 400$~GHz with a mean filter $Q_{filter}$ of $340 \pm 50$; instrument efficiency reached $\sim 8$~\%, indicating 4 times wider band coverage and 4 times higher sensitivity.
   The yield rate for MS filters exceeded 98~\%.
   Data loss caused by the sky-position chopper was limited to < 20~\% with no beam truncation.
   The estimated aperture efficiency from measured beam patterns agreed well with the designed value of approximately 70~\%.
   The telescope far-field beam patterns calculated from measured beam patterns also exhibited good agreement with design specifications.
   We also demonstrated validity of a new method of absolute frequency calibration using the data from beam pattern measurement.
   }
   {}


   \authorrunning{K. Karatsu et al.}
    \maketitle

%________________________________________________________________
\section{Introduction}
    
%                                  Two column figure
%__________________________________________________________________
   \begin{figure*}
   \centering
   \includegraphics[width=0.85\textwidth] {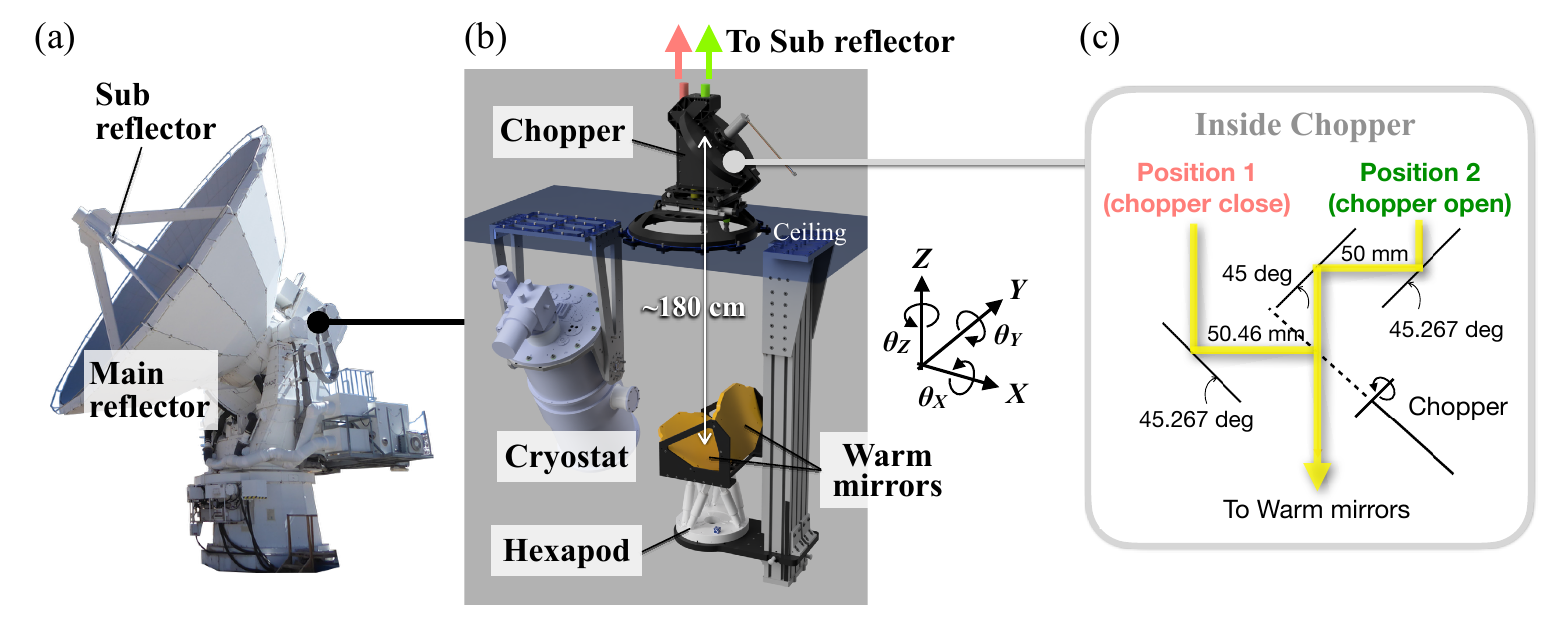}
      \caption{
   	(a) Picture of the ASTE telescope, whose main reflector has a diameter of 10~m.
	It is located at 4860~m altitude on Pampa la Bola, the Atacama desert, Chile.
	(b) Setup inside the ASTE cabin.
	The location of the sky-position chopper is $\sim180$~cm above from the warm mirrors.
	The red and green arrows correspond to two different beam positions (Position 1 and 2) shown in (c).
	(c) Schematic drawing of inside of the chopper.
	Two beams from Position 1 and 2 are separated by 100.46~mm, and reflected by mirrors with tilt angle of 45.267$^\circ$ to create 234~arcsec pointing separation in the sky.	
               }
              \label{Fig1}%
    \end{figure*}
%__________________________________________________________________

As millimeter (mm) / sub-millimeter (sub-mm) wave astronomy continues to evolve, there has been an increasing need for compact and wide-band spectrometers.
To meet this requirement, significant progress has been made in developing instruments using integrated superconducting spectrometer (ISS), such as DESHIMA  \citep[Deep Spectroscopic HIgh-z MApper,][]{Endo2019a, Taniguchi2022}, Superspec \citep{Karkare2020}, $\mu$-spec \citep{Mirzaei2020}, and SPT-SLIM \citep{Karkare2022}.
DESHIMA is a broadband spectrometer with a moderate frequency resolution ($R = 100 - 1000$).
By integrating superconducting resonant filters and kinetic inductance detectors \citep[KIDs,][]{Day2003}, DESHIMA embodies the concept of the ISS: The high multiplexing capability of KIDs enables an on-chip filterbank architecture, where each signal from an individual filter is read out by a corresponding KID.
This design allows for efficient and simultaneous measurement of multiple frequency channels across the (sub-)mm wave spectrum.
%\textcolor{red}{
Importantly, the ISS technology is scalable to an integrated field unit (IFU), by increasing the number of spatial pixels to $\gtrsim 100$.
%}

The first generation of DESHIMA \citep[DESHIMA~1.0,][]{Endo2019a}, covering frequency range of 332$-$377 GHz with a filter resolution ($Q_{filter}$) of  $\sim 300$ and an instrument efficiency of $\sim 2$~\%, was successfully tested on the Atacama Sub-millimeter Telescope Experiment (ASTE) telescope in 2017 \citep{Endo2019b}.

%\textcolor{red}{
In this paper we describe DESHIMA~2.0, the successor of DESHIMA~1.0.
DESHIMA~2.0 is an on-chip filterbank spectrometer with an instantaneous band of 200$-$400 GHz and a design spectral resolution ($\Delta F / F$) of 500.
In DESHIMA~2.0, we use 339 KIDs, read-out using a single back-end within a 2$-$4 GHz bandwidth.
Hence it covers $\sim 200$~GHz sky bandwidth with a 2~GHz readout bandwidth.
The separation between signal bandwidth and readout bandwidth makes the on-chip spectrometers very resource efficient compared to heterodyne/coherent systems.
%}

The primary science target of DESHIMA~2.0 is to study the cosmic history of star formation and galaxy evolution by detecting molecular and atomic emission lines from bright dusty star-forming galaxies at high redshifts \citep{Rybak2022}.
%\textcolor{red}{
The moderate frequency resolution of 500 is designed to match the typical line velocity width of the most massive dusty star-forming galaxies \citep[$\Delta \upsilon = 600$~km/s,][]{Carilli2006}.
%}
The broadband feature also enables us to probe dynamics of galaxy clusters through spectroscopic observations of the Sunyaev-Zel'dovich effect \citep{SZ1970, SZ1980}.

%DESHIMA~2.0 is the upgrade of DESHIMA~1.0, offering broader frequency coverage and enhanced sensitivity by combining an ultra-wideband leaky-wave antenna \citep[][Dabironezare et al. in prep.]{Neto2010a, Neto2010b, Hahnle2020} and an NbTiN superconducting micro-strip (MS) filter bank spectrometer \citep{Laguna2021, Thoen2022} with low-loss a-SiC:H dielectric \citep{Buijtendorp2022, Buijtendorp2025}. 
Compared to DESHIMA~1.0, DESHIMA~2.0 offers broader frequency coverage and enhanced sensitivity by combining an ultra-wideband leaky-wave antenna \citep[][Dabironezare et al. in prep.]{Neto2010a, Neto2010b, Hahnle2020} and an NbTiN superconducting micro-strip (MS) filter bank spectrometer \citep{Laguna2021, Thoen2022} with low-loss a-SiC:H dielectric \citep{Buijtendorp2022, Buijtendorp2025}. 
The DESHIMA~2.0 instrument also features a fast sky-position chopper that enables more efficient removal of atmospheric fluctuations compared to its predecessor, resulting in longer on-source time fractions.
Finally, it features a higher transmission optics, in particular the anti-reflection (AR) coated cryostat window and infrared filters.
%Table~\ref{Table1} summarizes the specification of DESHIMA~2.0 with the specification of DESHIMA~1.0 as comparison.
%\textcolor{red}{
The DESHIMA~2.0 instrument was deployed to the ASTE telescope in 2023, and we carried out one year long on-sky campaign \citep[][Endo et al. in prep.]{Moerman2025}.
%}

%In this paper we describe the design of the DESHIMA~2.0 instrument and the full characterization of the instrument in the laboratory.
The design of the DESHIMA~2.0 instrument and the full characterization of the instrument in the laboratory are described in this paper.
The paper structure is as follows: We explain the specification of the DESHIMA~2.0 instrument in Sec.~\ref{deshima2} together with its cryogenic configuration.
Sec.~\ref{deshima2chip} describes details of the ISS chip of DESHIMA~2.0.
Then, in Sec.~\ref{labsetup}, we discuss the laboratory setup and experiments we performed to characterize the instrument.
In Sec.~\ref{freqresponse}, \ref{sensitivity}, and \ref{beampattern} we show the results of the on-chip filter characterization, the instrument sensitivity, and the beam characteristics of the DESHIMA~2.0 instrument, respectively.
We show in Sec.~\ref{freqcalibration} the result of the absolute frequency calibration of the instrument done by comparing filter responses with output frequencies from the harmonic mixer, which was used in the beam pattern measurements.
Finally, Sec.~\ref{conclusion} is the conclusion.
  
%%                                  One column table
%%__________________________________________________________________
%   \begin{table}
%   	\caption{
%		Specification of the DESHIMA instrument.
%		     }
%		     \label{Table1}
%	\scalebox{0.9}[1.0]{
%	\begin{tabular}{l|c|c}
%	  & DESHIMA~1.0 & DESHIMA~2.0 \\
%	  \hline \hline
%	  Freq. range (GHz) & 332 - 376 & 200 - 400 \\
%	  HPBW (arcsec) & 22 & 32 - 17 \\
%	  Freq. spacing ($Q_{space}$) & 380 (design) & 500 (design) \\
%	  $Q_{filter}$ & 300 (measured) & 500 (design) \\
%	  Num. of spectral channels  & 49 & 339 \\
%	  Num. of spatial pixels & 1 & 1 \\
%	  %Num. of polarization & 1 & 1 \\
%	  %\hline \hline
%	  Inst. efficiency ($\%$) & 2 & $> 8$ \\
%	  %In band efficiency  ($\%$)  & & \\
%	  %Usable spectrum fraction  ($\%$)  & & \\
%	  Chopping & \begin{tabular}[c]{@{}l@{}} 10 Hz between \\ 300 K and sky \end{tabular}  & \begin{tabular}[c]{@{}l@{}} 10 Hz between \\two sky positions \end{tabular} \\
%	  Beam separation (arcsec) & -- & 234 \\
%	  Hold time (hour) & $> 20$ & $> 20$ \\
%	  %\hline
%	\end{tabular}
%	}
%   \end{table}
%%__________________________________________________________________

%__________________________________________________________________
\section{Methods}

%__________________________________________________________________
\subsection{The DESHIMA 2.0 instrument} \label{deshima2}

%                                  Two column figure
%__________________________________________________________________
   %\begin{figure}
   \begin{figure*}
   \centering
   \includegraphics[width=0.85\textwidth] {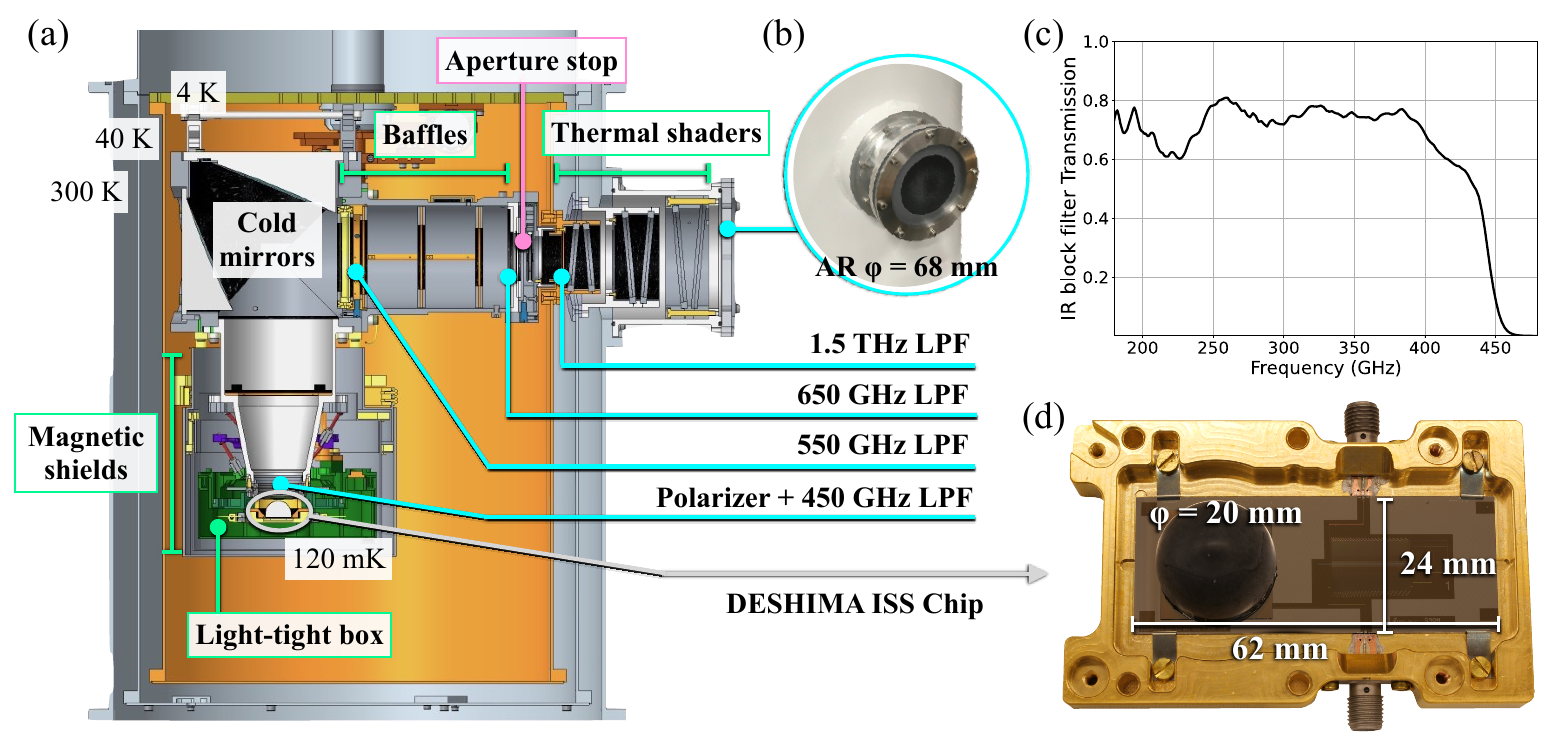}
   \caption{
   	(a) Cross section and overview of the DESHIMA cryostat with explanations of components.
	(b) Picture of Si vacuum window with laser-ablated AR structures.
	(c) Total transmittance of Si window, thermal shaders, and IR blocking filters along the cryogenic optical throughput.
	Four kinds of IR blocking filters: 1.5~THz low-pass filter (LPF), 650~GHz LPF, 550~GHz LPF, and 450~GHz LPF are installed in the optical throughput.
	(d) Picture of the DESHIMA 2.0 spectrometer (ISS) chip.
   		}
              \label{Fig2}%
   \end{figure*}          
   % \end{figure}          
%__________________________________________________________________

The DESHIMA~2.0 instrument is designed to fit in the ASTE telescope (Fig.~\ref{Fig1} (a), (b)).
Figure~\ref{Fig1} (b) shows the interior of the ASTE cabin with the DESHIMA~2.0 instrument.
It consists of three main components:
1) a sky-position chopper with a motor-controlled stage;
2) room temperature warm mirrors supported by a hexapod;
and 3) the cryostat housing the DESHIMA ISS chip together with cryogenic optics.
All of these are attached to the ceiling of the ASTE cabin, thereby creating a stable relative position between these three components, as well as a first order absolute alignment.
Precise optical alignment between the ISS in the cryostat, the telescope optics and the beam modulator is done with the motor-controlled stage of the sky position chopper and the hexapod, which is discussed in detail in \citet{Moerman2024}.
%The detail of alignment procedure is described in \citet{Moerman2024}.

The sky-position chopper consists of three planar mirrors and a wheel. 
Figure~\ref{Fig1} (c) displays schematic drawing of the inside of the chopper.
The optical pass through the chopper changes by rotating the wheel (denoted as Position~1 and 2 in Fig.~\ref{Fig1} (c)).
The chopper wheel rotates at a frequency of 5~Hz, resulting in a chopping frequency of 10 Hz with close to 50~\% duty cycle as the wheel has two slots, each of which occupies a quarter of the surface area of the wheel.
In order to monitor the phase (slot or open) of the wheel, a diode sensor is installed inside the chopper module.
The chopping frequency is set not only to remove atmospheric fluctuation (order of $<$~1~Hz), but also to suppresses the 1/f noise due to the two-level system (TLS) noise in the KIDs \citep{Gao2008a}.	
The chopper is positioned on a motor-controlled stage, which travel range is $\pm 25$~mm in X/Y direction with 10~$\mu$m accuracy (the definition of the coordinates is shown in Fig.~\ref{Fig1} (b)).
Two beams from Position~1 and 2 illuminate the sub-reflector of the ASTE telescope with a separating distance of $\sim100$~mm, resulting in a $\pm117$~arcsec pointing offset along the azimuthal direction with respect to broadside.
The azimuthal offset is chosen to be large enough compared with the beam size of the ASTE telescope over the DESHIMA~2.0 frequency band of $200-400$~GHz, and, at the same time, to be small enough to see the same patch of atmosphere for effective removal of the atmospheric fluctuation.

The two warm mirrors consists of a hyperboloidal mirror and an ellipsoidal mirror that are placed in a modified Dragonian configuration \citep{Dragone1978}.
The mirrors are inside a single rigid support structure, which is is mounted onto a six-axis motorized hexapod.
The travel range of the hexapod is $\pm 50$ (25)~mm in X/Y (Z) direction with 4~$\mu$m accuracy and rotation range is $\pm 15^{\circ}~(30^{\circ})$  in $\theta_X/\theta_Y$ ($\theta_Z$) with 7.5~$\mu$rad accuracy.
%The purpose of the warm mirrors is to couple the cryogenic optics inside the cryostat to the ASTE Cassegrain optics.

Figure~\ref{Fig2} (a) displays an overview of the DESHIMA cryostat, which is the same two-stage adiabatic demagnetization refrigerator (ADR, Entropy GmbH) as used for DESHIMA~1.0 \citep{Endo2019a}.
The cryostat has a 6~mm thick silicon (Si) vacuum window, which is AR coated with a laser-ablated sub-wavelength structure over its central 68~mm diameter (Fig.~\ref{Fig2} (b)).
The transmittance of the window is more than 95~\% over the frequency band of DESHIMA~2.0 (Takaku et al. in prep.).
Along the optical path inside the cryostat, thermal shaders and infrared (IR) blocking filters are installed, as well as baffles coated by a radiation absorber \citep{Klaassen2001} to limit large angular stray radiation.
Figure~\ref{Fig2} (c) is the total transmittance of the combination of Si window, thermal shaders, and IR blocking filters as a function of frequency, showing an average transmission of 70~\% is achieved over the frequency band of DESHIMA~2.0.
This is higher than the filters and window for DESHIMA~1.0, which was $\sim 40$~\% \citep{Endo2019a}.
The cold mirrors consist of a parabolic relay with two off-axis paraboloid reflectors that are mounted on the 4 K stage as one unit.
The cryogenic optics couple the radiation entering the cryostat to the ISS chip that is located at 120~mK stage inside the light-tight box \citep{Baselmans2012}.
The light-tight box, which holds the detector chip, is thermally anchored to the cold stage of the ADR, and is operated at 120~mK.
It is surrounded by dual layer magnetic shield of Nb and Cryoperm, cooled to 4~K.
The thermal isolation between the 4~K optics box and the 120~mK light tight box is achieved using a thermal-mechanical assembly identical to the one used by DESHIMA~1.0, which is based upon vespel rods and in-house made thermal straps.
It uses an intermediate thermal stage, connected to the 0.8~K stage of the ADR to reduce the heat load on 120~mK, which is also used to thermalize the coax cables used for the readout. 
Figure~\ref{Fig2} (d) is a picture of the DESHIMA ISS chip with its holder.
It is equipped with a Si lens of 20 mm diameter, which is AR coated with a $\sim120~\mu$m thickness of a mixture of Stycast 2850FT and 1266A to make reflective index n = 1.84 \citep{Suzuki2012, Nitta2014}.
The lens is placed at a distance of 10~$\mu$m from the leaky-wave antenna feed that connects to the filterbank.
The lens-antenna, cold and warm optics provide a highly efficient radiation coupling over a the full DESHIMA~2.0 band, which is described in detail in (Dabironezare et al. in prep.).
In front of the lens, a polarizer is also mounted to select the polarization of the incoming radiation.
The detailed design of the ISS chip is described in the next section (Sec.~\ref{deshima2chip}).

The ISS chip is connected to the cryogenic readout chain by a pair of coax lines.
A low noise amplifier with $+26$~dB gain and noise temperature of $\sim2$~K (Low Noise Factory, LNF-LNC1.5\_6A s/n 750B) is located at the 4~K stage in the output line of the readout chain.
The readout signal is further amplified at right after the exit of the cryostat by an amplifier with +20~dB gain.
The readout chain is connected to a room temperature, frequency-division multiplexing readout system \citep{Rantwijk2016} to read out the response of all KIDs simultaneously.
The sampling speed of the readout system could be selected from $\sim160$~Hz or $\sim1.3$~kHz, and we used both settings depending on the requirements from experiments.

The total loading power, including the power from the optical throughput, the mechanical support structure and the readout chain, was measured to be $\sim2~\mu$W at the coldest stage.
With the operation temperature of 120~mK, we measured a hold time of the ADR of more than 24~hours.

%__________________________________________________________________
\subsection{The DESHIMA 2.0 ISS chip} \label{deshima2chip}

%                                  Two column figure
%__________________________________________________________________
   \begin{figure*}
   \centering
   \includegraphics[width=0.95\textwidth] {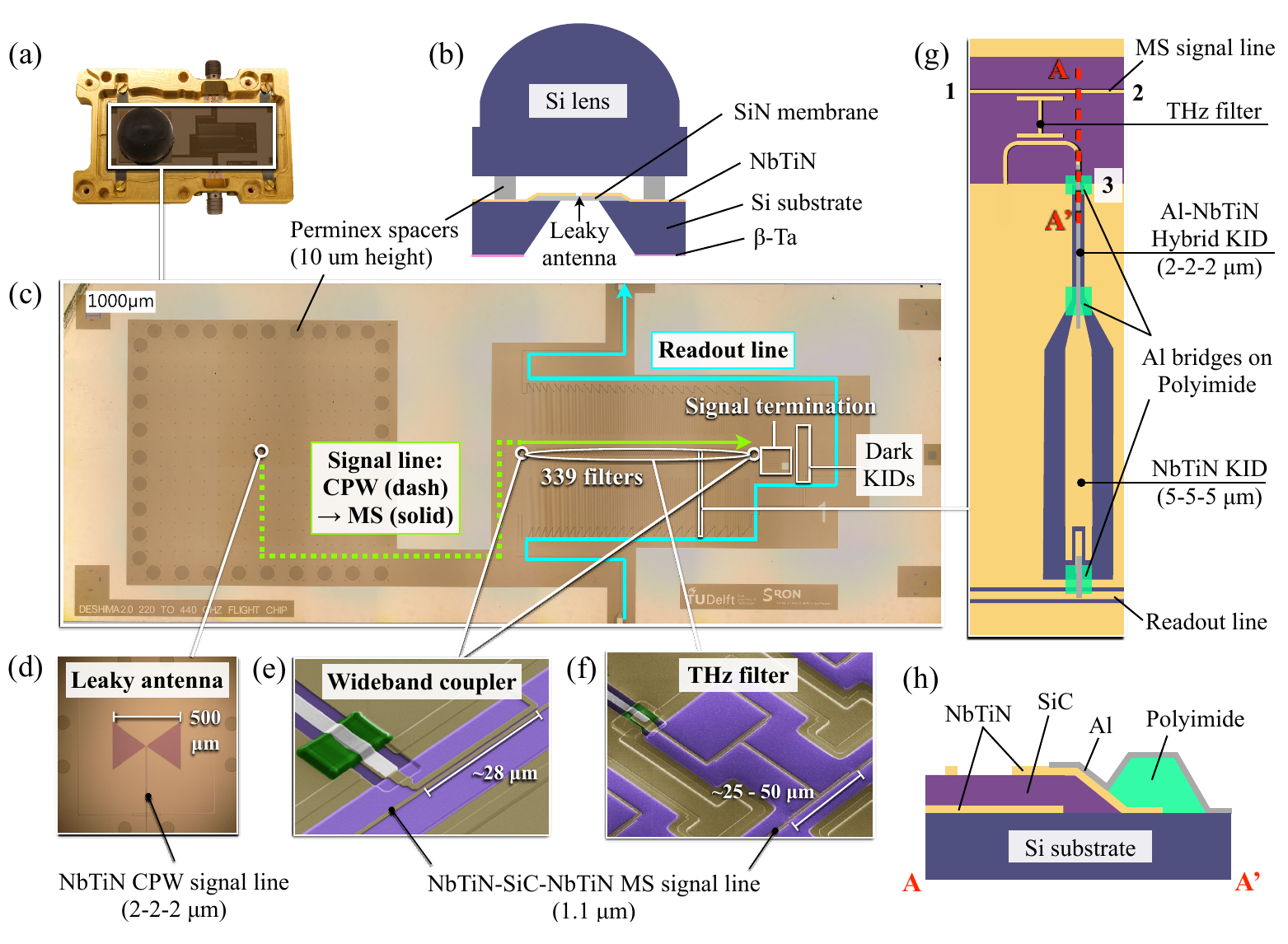}
   \caption{
   	(a) Photograph of the DESHIMA~2.0 spectrometer chip with its holder.
	The white rectangle indicates the spectrometer chip, which is close-up in (c).
	(b) Cross section around the Si lens and the antenna (not to scale).
	(c) Micrograph of the spectrometer chip.
	The location of the leaky-wave antenna, the signal line, the wideband couplers, the THz filters (filterbank), the signal termination, dark KIDs and the readout line are indicated.
	The Perminex spacers of $10~\mu$m height around the antenna are also shown.
	(d) Zoomed-in micrograph of the leaky-wave antenna on the SiN membrane.
	The antenna width is $500~\mu$m, and it has a CPW based quarter-wave impedance matching structure that eventually connects to the NbTiN CPW signal line with a geometry of $2-2-2~\mu$m.
	(e) Colored scanning electron microscope (SEM) image of the wideband coupler.
	The coupler length  is $\sim28~\mu$m, weakly coupled to the MS signal line of $1.1~\mu$m width.
	The coupling strength is designed to be $-29 \pm 1$~dB over the frequency range of DESHIMA 2.0.
	(f) Colored SEM image of the THz filter (image taken from \citet{Thoen2022}).
	The filter coupler length varies from $\sim25 - 50~\mu$m depending on the resonance frequency of the filter.
	The distance between the filter and the signal line is designed to be 300~nm.
	In the filterbank, the spectral channels are placed in decreasing order of frequency from 400~GHz to 200~GHz.
	(g) Schematic drawing of a pair of MS THz filter and KID, showing how each component is connected or coupled.
	KID consists of a short-ended NbTiN MS line with a width of 1.5~$\mu$m that couples to the filter, NbTiN-Al hybrid CPW line with a geometry of $2-2-2~\mu$m, and open-ended NbTiN CPW line with a geometry of $5-5-5~\mu$m.
	Al bridges are used to electrically connect the NbTiN MS line, the NbTiN-Al hybrid CPW line and the NbTiN CPW line of the KID.
	The open-end of the KID is capacitively coupled to the CPW readout line with an Al bridge.
	Not displayed in the drawing, Al bridges are also used to balance the ground planes of the readout CPW line by placing them in a random spacing.
	(h) Cross section of the connection between the MS line and the Al line of the hybrid CPW (not to scale). The corresponding location is indicated by the red dashed line in (g).
		}
              \label{Fig3}%
    \end{figure*}
%__________________________________________________________________    

The size of the DESHIMA~2.0 ISS chip is 62~mm $\times$ 24~mm as shown in Fig.~\ref{Fig2} (d).
It is fabricated from a 350~$\mathrm{\mu}$m FZ $<100>$ Si wafer with $\rho\;>$ 10~$\mathrm{k\Omega cm}$ coated on both sides by a 1~$\mathrm{\mu}$m LPCVD (Low-Pressure Chemical Vapor Deposition)  Si-rich SiN layer with a (tensile) stress $\sigma\approx$~$+150$~MPa.
This layer is removed everywhere using Reactive Ion Etching (RIE) except on the wafer front side where the antenna will be located.
We use a plasma with 13.5~sccm Ar$^+$ and 25~sccm O$\mathrm{_2}$ at 50~W and 5 mTorr, followed by a 90 second pure $\mathrm{O_2}$ descum, to create a sloped edge on the SiN for later ease of step coverage. \\
The second layer is the NbTiN ground, deposited using shuttled reactive magnetron sputtering from a NbTi target in a Argon-Nitrogen plasma to deposit a very flat 200 nm NbTiN layer \citep{Thoen2016} with a $T_c = 14.7$~K that is also patterned using the same sloped edge RIE process to create the ground plane of the MS filterbank. \\
The third layer is the filterbank dielectric, fabricated from a 500~nm thick a-SiC:H deposited at 400~$^\circ$C using plasma enhanced chemical vapor deposition, (PECVD) \citep{Buijtendorp2022}.
We again use the sloped edge RIE process to pattern the a-SiC:H, which is almost everywhere smaller than the NbTiN ground plane to create a good galvanic connection with the next NbTiN top layer, as shown in panels (e) and (f) of Fig.\ref{Fig3}.
However, where the KIDs connect to the filters only the KID central line needs to be connected to the NbTiN top layer, so here the a-H:SiC is patterned to overlap the NbTiN ground, as shown in the  panels (g) and (h) of Fig.\ref{Fig3}. \\
The fourth layer is defined from a 207 nm NbTiN layer, identical to the ground plane layer.
We used a dual e-beam/UV exposure of ma-N1405 negative tone resist, which is sensitive to both the 365~nm UV radiation from our UV mask aligner as well as to the 100~keV electrons from our electron beam pattern generator (EBPG).
The mask aligner is used to expose the large ground planes and coarse structures, the EBPG is used to define the signal line and the KID couplers.
The details of this mix-and-match fabrication are described in \citep{Thoen2022}.
After development the NbTiN is etched in the RIE using a recipe with less oxygen to create steeper slopes and better edge definition: We used a 25~sccm Ar - 13.5~sccm O$\mathrm{_2}$ plasma at 50~W and 5~mTorr, followed by a 90 second pure $\mathrm{O_2}$ descum. \\
The fifth layer consists of a 500~nm thick polyimide, which is photosensitive and patterned into small rectangles to support the aluminum (Al) of the bridges over the readout line and to the KID couplers.\\
The sixth layer is a 35~nm thick Aluminium (Al) ($T_c = 1.2$~K), which is used as radiation absorber in the the NbTiN-Al hybrid KIDs \citep{Janssen2013, Baselmans2017} and for the Al bridges.
This Al is sputter deposited using shuttling with the same target dimensions and method as the NbTiN layers \citep{Thoen2016} to create a very flat Al layer.
This layer is patterned in two steps: First we do a normal UV exposure and wet etching to define the bridges and the Al in the KIDs, with the latter a few micron wider than their final dimension.
In the next step, we use PMMA, EBPG patterning and wet etching to very accurately define the final Al linewidth.
The use of very flat NbTiN and Al layers, as well as EBPG patterning of the filters and KID Al, is key in reducing frequency scatter of both the KIDs and filters as we will see later. \\
The seventh layer is 10~$\mu$m Perminex (KAYAKU PermiNex 1010), deposited using spin coating, UV exposure and developing.
This material is used to define the vacuum gap at the leaky-wave antenna and glue the Si spacer-Si lens to the chip, similar as in \citep{Baselmans2022}. % that serves as spacers between the Si lens and the leaky-wave antenna.
When all these layers are patterned, we use KOH etching to remove the Si behind the SiN with the leaky-wave antenna feed to create a free standing membrane. \\
The final layer is a 40~nm $\beta$-phase Ta ($T_c \sim 0.7$~K) deposited on the backside of the chip as an absorbing mesh for stray-light control \citep{Yates2017}.
This layer is created using sputter deposition, UV patterning and dry etching. We use careful resist placement and spinning to prevent resist filling up the holes in the Si at the positions of the SiN membranes.

Figure~\ref{Fig3} (a) is the same picture as Fig.~\ref{Fig2} (d), but with an indication of the DESHIMA ISS chip, which micrograph is shown in Fig.~\ref{Fig3} (c).
Figure~\ref{Fig3} (b) displays a schematic drawing of cross-section around the Si lens and the antenna (not to scale).
The Si lens is fixed and supported by 10~$\mu$m high Perminex pillars on the chip.
Alignment between the Si lens and the antenna was done by using the alignment marker patterned at the center of the Si lens and the center of the antenna.
The linearly polarized incoming radiation coming through the Si lens is received by the broadband leaky-wave antenna on the SiN membrane (Fig.~\ref{Fig3} (d)).
The received signal is guided to the filterbank through the coplanar waveguide (CPW) signal line that is converted to a MS line right before entering to the filterbank as indicated by the light-green lines in Fig.~\ref{Fig3} (c) (CPW part: dashed line, MS part: solid line).
The signal is then sorted to 339 spectral channels over the $\sim200-400$~GHz frequency range of DESHIMA~2.0 by 339 narrow-bandpass half-wavelength MS filters \citep[called THz filter hereafter,][]{Laguna2021} in the filterbank.
The separation of the neighbouring filters of the resonance frequencies of $F_{filter}^i$ and $F_{filter}^{i+1}$ ($i$ denotes $i$-th filter) is designed as $\lambda_{eff}^i/4 = c/ (4 F_{filter}^i \sqrt{\varepsilon_{eff}})$ where $c$ is the speed on light in free space, and $\varepsilon_{eff} \sim 30$ describes the effective phase velocity of the MS line, $\upsilon_{ph} = c / \sqrt{ \varepsilon_{eff} },$ including both the permitivity of the dielectric and the kinetic inductance of the NbTiN. %is the effective permitivity of the MS signal line.
As a reference for the coming signal power, four wideband couplers are placed, two before and other two after the filterbank, that are weakly coupled to the signal line.
The separation of the wideband couplers is $\lambda_{eff}/4$ at 300~GHz, the center frequency of the DESHIMA~2.0 band.
Figure~\ref{Fig3} (e) and (f) show colored scanning electron microscope (SEM) images of the wideband coupler and the THz filter, respectively.
Figure~\ref{Fig3} (g) displays a schematic drawing of a pair of THz filter and corresponding KID.
The MS filter is coupled to the MS signal line on one side, and to a KID on the other side.
The KID is a quarter wavelength hybrid resonator \citep{Janssen2013}, consisting of a short-ended NbTiN MS line that couples to the filter, NbTiN-Al hybrid CPW line, and open-ended NbTiN CPW line.
%We use a polyimide support with an Al layer on top to form bridges to electrically connect the NbTiN MS line, the NbTiN-Al hybrid CPW line and the NbTiN CPW line.
We use a polyimide support wherever we need to galvanically connect Al to NbTiN to prevent yield issue due to trenching.
The open-end of the KID is capacitively coupled to the CPW readout line via a coupling structure that is galvanically connected to the readout line by such an Al-polyimide bridge.
Figure~\ref{Fig3} (h) shows a schematic drawing of cross section of the connection between the MS line and the Al line of the hybrid CPW (not to scale, the corresponding location is indicated by the red dashed line A-A' in Fig.~\ref{Fig3} (g)).
After the filterbank, the signal line is terminated by a CPW line with an Al center line to absorb the remaining signal, which prevents reflection of power that is not absorbed by the filter channels.
In the chip, four dark KIDs, not connected to filters or coupled to the signal line, are placed at the location away from the signal line to measure the power level of stray light or surface waves in the substrate.
The locations of both the signal termination and dark KIDs are also shown in Fig.~\ref{Fig3} (c).
All KIDs are coupled to one readout line, and the microwave resonance frequencies are designed to be distributed within a $4-6$~GHz range with a step of $\Delta f_r = 4.5$~MHz.
The in-band power coupled to an individual MS filter is absorbed in the Al of the hybrid CPW, changing the surface impedance and hence the resonant frequency of the KID.
The shift of the resonance frequency is readout out as a change in the transmission amplitude and phase of a microwave tone in the readout line.

%__________________________________________________________________
\subsection{Setup and experiments in the laboratory}  \label{labsetup}

%                                  One column figure
%__________________________________________________________________
   \begin{figure}
   \centering
   \includegraphics[width=0.45\textwidth] {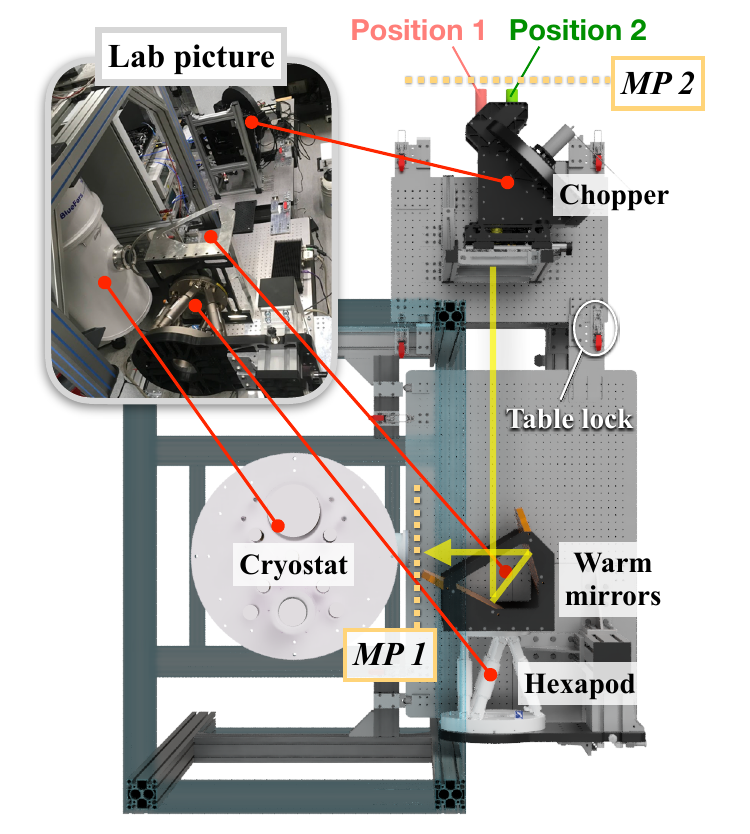}
      \caption{
      	Rendered CAD drawing of the laboratory setup with the inset picture of the laboratory setup.
	The red lines correspond each components of the instrument in the lab picture to the rendered drawing.
	The optical path from the sky chopper to the cryostat is shown as the yellow arrow.
	The definition of two measurement planes (MP1 and 2) are indicated as yellow dashed lines.
              }
         \label{Fig4}
   \end{figure}

In order to characterize the DESHIMA~2.0 instrument, we created a $1:1$ copy of the telescope cabin in our laboratory where the three main components (sky chopper, warm mirrors, cryostat) of the instrument were located at the same relative positions as inside the telescope cabin.
For practical reasons we rotated the entire system by 90 degree so that the warm mirrors and the sky chopper were positioned horizontally as shown in  Fig~\ref{Fig4}.
We placed the warm mirrors and chopper each on a separate small optical table and used mechanical locks to create a rough alignment which was extremely reproducible.
This allowed for efficient sub-system testing by removing 1 or 2 optical tables.
All distances were identical to the telescope cabin situation.
Once the relative positions were roughly fixed, precise positioning and alignment could be done with the motor-controlled stage attached to the sky chopper and the hexapod attached to the warm mirrors, an identical strategy as in the telescope cabin.

We carried out a series of experiments to characterise the DESHIMA 2.0 instrument as follows:
\begin{description}
 \item [\textbf{Step~1:}] We measure the frequency response of all THz filters at the measurement plane (MP)~1, in front of the cryostat window and indicated in Fig.~\ref{Fig4}.
 This was done using a continuous wave (CW) frequency source.
 \item [\textbf{Step~2:}] We measure the beam patterns at MP~1 to check the phase and amplitude (PA) beam patterns in front of the cryostat window.
 \item [\textbf{Step~3:}] After aligning the warm mirrors and the chopper positions with respect to the cryostat, we measure the beam patterns at MP~2, which is after the chopper module as indicated in Fig.~\ref{Fig4}.
The beam patterns of both optical paths (Position~1 and 2) of the chopper were measured separately by fixing the chopper position at either position.
 \item [\textbf{Step~4:}] We measure the instrument noise equivalent power ($NEP_{inst}$) and optical coupling efficiency ($\eta_{inst}$) at MP~2 to evaluate optical sensitivity of the instrument.
This measurement could not be done without a proper alignment in Step~3.
We took data in two configurations: $T_{cold}$ at Position~1 and $T_{room}$ at Position~2, and vice versa, in order to characterise both optical paths separately.
\end{description}
  
The alignment at Step~3 was done by simulating the warm mirrors and hexapod and performing a ray tracing analysis through the warm mirrors to calculate the optimal hexapod configuration, using the the beam patterns obtained at Step~2 as input. %to take into account the relative position of the cryostat (cryogenic optics) and the cryostat supporting frame.
The ray-tracing calculations and simulations of the warm optics are done using the optical simulation software PyPO \citep{Moerman2023}.
The detail of alignment procedure (Step~2 and 3) is described in \citet{Moerman2024}.
Step~3 also involved some iterations to get correct information of the relative position between the optical components and the measurement planes (MP~1 and 2) in the real setup in order to properly implement MP~1 and 2 to the simulation.
This was not only for improving the alignment quality, but also essential to propagate the measured beam patters through a model of ASTE using a physical optics calculation, which is also supported by PyPO, to estimate expected performance of the instrument at the telescope.

In above experiments, the changes in the transmission amplitude and phase of microwave tones, tuned to each KID, were read out when taking time domain data.
We also read out several microwave tones that were off from KID frequencies (called blind tones) to measure the readout electronics noise.
In the data analysis, we converted the phase change ($\delta \theta$) to the microwave frequency response ($x \equiv (f-f_r)/f_r = \delta f/f_r$, where $f_r$ represents resonance frequency of the KID) by using KID phase versus frequency dependence curve obtained from so-called the local sweep measurement \citep{Bisigello2016}.
%by using the relation of $x = - \delta \theta / 4 Q_r$ \citep{Gao2008b}, where $Q_r$ is quality factor of a KID obtained from so-called the local sweep measurement.
In the local sweep measurement, the frequency of the microwave tones were locally swept in a range of 2~MHz with 10~kHz steps by sweeping the frequency of the local oscillator used in the readout system, so that we could measure the complex $S_{21}$ transmission parameters of all KIDs simultaneously and extract KID parameters such as $f_r$ and $Q_r$ by a Lorentzian fit of the measured $S_{21}$ \citep{Gao2008b}.
The local sweep data was obtained every time before the measurement of time domain data.
%%

%%__________________________________________________________________
%\section{Results and discussions}
%
%%__________________________________________________________________
%\subsection{THz filter characteristics} \label{freqresponse}

%                                  Two column figure
%__________________________________________________________________
   \begin{figure*}
   \centering
   \includegraphics[width=0.95\textwidth] {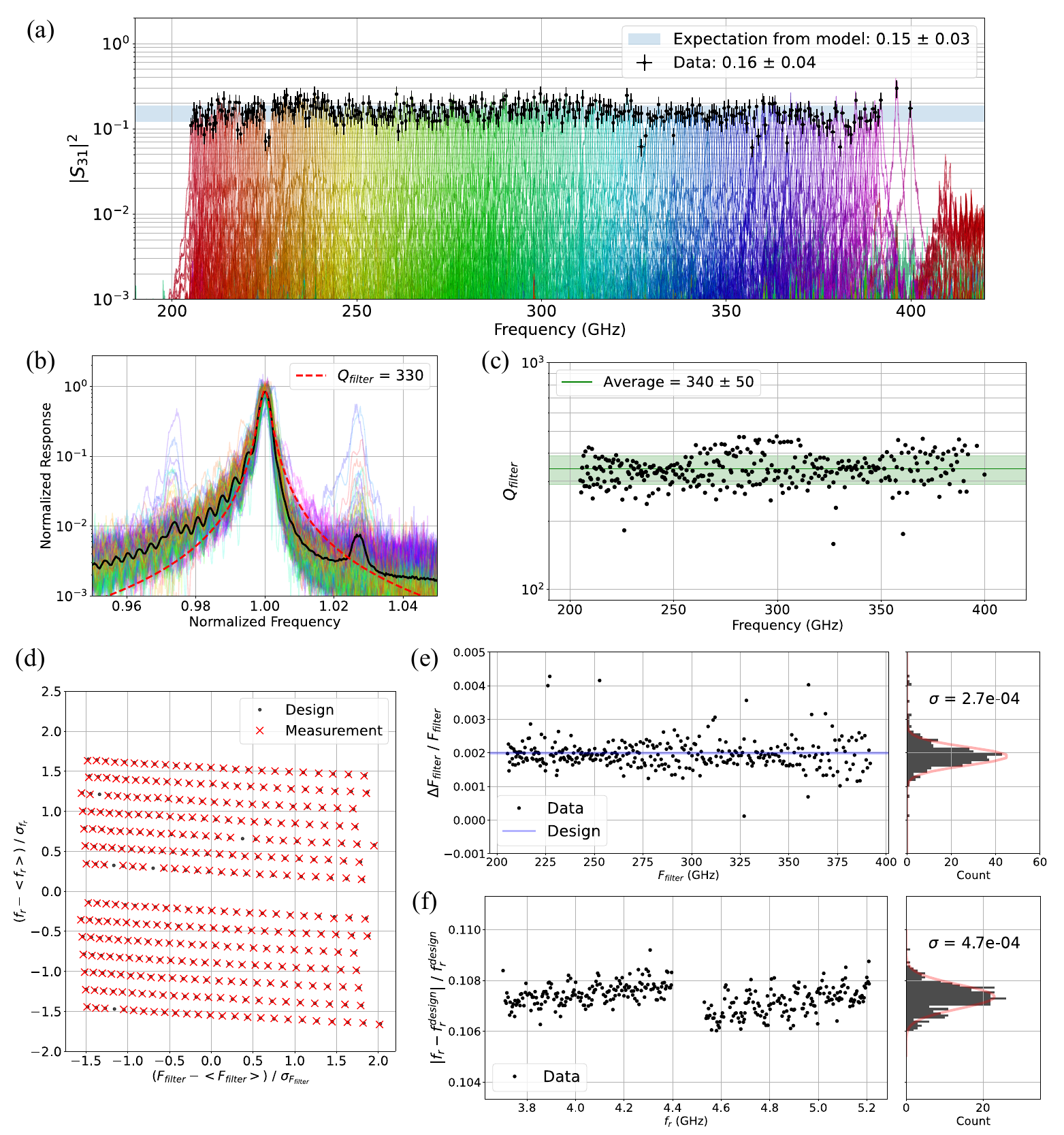}
   \caption{
   (a) Measured response of the DESHIMA~2.0 spectral channels, $|S_{31}|^2$.
   Because they are half-wavelength filters, the second resonances from the lower frequency range (around $200-220$~GHz) are also visible above 400~GHz.
   The black points represent $|S_{31}|^2_{max}$ with error bars, which are combination of errors from a Lorentzian fit and the design value of $\eta_{wb, bf}$.
   The blue shaded box shows the expected $|S_{31}|^2_{max}$ from a numerical filterbank model.
   %(b) Zoom-in view of the response of one THz filter KID around the middle of the band (black curve), while the red curve represents the results from a Lorentzain fit.
   %The obtained parameters from the fit are listed in the legend.
   (b) THz filter responses normalized in the frequency domain (colored curves) and averaged shape (black curve).
   The red dashed curve shows the result from a Lorentzian fitting to the averaged shape.
   The obtained $Q_{filter}$ from the fit is $\sim 330$.
   From the peaks at the normalized frequency of around 0.975 and 1.025 of the averaged shape, we can also estimate a cross-talk level of $\sim 1$~\% that is mainly caused by the cross talk by neighbour KIDs in the microwave frequency domain.
   (c) The obtained $Q_{filter}$ from the fitting for all THz filters as a function of $F_{filter}$ (black points).
   The green solid line and shaded area displays the averaged $Q_{filter}$ and its standard deviation, respectively.
   (d) The 2-dimensional distribution of normalized $(F_{filter}, f_r)$.
   The black points are design, and the red crosses show the measured data, showing good correspondence between them.
   (e) The black points in the left panel shows calculated frequency spacing of the THz filters ($\Delta F_{filter} / F_{filter} = (F_{filter}^{i+1} - F_{filter}^{i}) / F_{filter}^i$) as a function of $F_{filter}$, while the blue line indicates the design value of 0.02 ($= 1/500$).
   The right panel displays the distribution of the frequency spacing with the value of standard deviation ($\sigma$) obtained from a Gaussian fit.
   The result of the fitting is also shown as the red curve in the right panel.
   (f) The black points in the left panel shows calculated frequency-scattering of KIDs ($|f_r^{measured} - f_r^{design}|/f_r^{design}$) as a function of $f_r$.
   The right panel displays the distribution of the frequency-scattering with the value of standard deviation ($\sigma$) obtained from a Gaussian fit.
   The result of the fitting is also shown as the red curve in the right panel.
   }
              \label{Fig5}%
    \end{figure*}

%__________________________________________________________________
\section{Results and discussions}

In this section, we first present the THz filter characteristics from Step~1 in Sec.~\ref{freqresponse}.
We then show the optical sensitivity of the instrument from Step~4 in Sec.~\ref{sensitivity}.
In Sec.~\ref{beampattern}, we describe the expected beam characteristics at the telescope calculated using PyPO by propagating the measure PA beam patterns from Step~3.
Finally, the result of the absolute frequency calibration using the results from Step~1 and 3 is presented in Sec.~\ref{freqcalibration}.

%__________________________________________________________________
\subsection{THz filter characteristics} \label{freqresponse}

For measurement of frequency response of THz filter, we used a photo-mixing THz CW source mounted on a third optical table (not shown in Fig.~\ref{Fig4}), directly in front of the cryostat window, that can emit signal in the frequency range of 100~GHz to 1.2~THz with a relative frequency accuracy of 20~MHz.
 The signal power of the CW source was modulated by 12~Hz and so modulated the detector response in time domain.
 We carried out the experiment with a sampling frequency of $\sim 160$~Hz, which is large enough to trace the 12~Hz modulation.
 By sweeping the frequency of the CW source from 180~GHz to 460~GHz in 50~MHz step, we measured the response of all KIDs simultaneously, from which we characterised THz frequency response of each filter.
This allowed us to correlate the KID readout frequency to an individual THz filter frequency, and compare both to the designed values.
 More details of the experiment are described in \citet{Endo2019a}.
 Note that the absolute frequency accuracy of the CW source was 2~GHz, so another measurement was required for absolute frequency calibration of the THz filter frequency, which is discussed in Sec.~\ref{freqcalibration}. 

Figure~\ref{Fig5} (a) shows the frequency response of the DESHIMA~2.0 ISS, which is defined as the response of the KID at filter $i$, caused by the power at Port~3 in Fig.~\ref{Fig3} (g) entering the KID due to a THz signal at Port~1 (see Fig.~\ref{Fig3} (g)) entering the filter bank.
This is defined as $S_{31}$, given by \citep{Laguna2021} :
\begin{eqnarray}
\centering
% \big| S_{31}(F) \big|^2 \sim \frac{ R_{filter} (F) }{ R_{wb, bf}(F) }~\big| S_{31}^{wb, bf} \big|^2
 \big| S_{31}^i (F) \big|^2 &=& \frac{ x_{filter}^i (F)~( dx_{filter}^i / dP_{abs}^i )^{-1} }{ x_{wb, bf}(F)~(dx_{wb, bf}/dP_{abs}^{wb,bf})^{-1} }~\eta_{wb, bf} \label{Eq1} \\
 &\sim& \frac{ x_{filter}^i (F) }{ x_{wb, bf}(F) }~\eta_{wb, bf} \label{Eq2}
\end{eqnarray}
where $i$ represents $i$-th filter, $x_{filter}^i (F)$ and $x_{wb,bf} (F)$ are the change in KID resonant frequency  ($x = \delta f/f_r$) due to the applied CW signal at THz filter $i$ and wideband KID, respectively.
Note that the linearity of the detector respsonse was checked by repeating the experiment at a lower CW source power, which gave the same result.
$dx/dP_{abs}$ is the KID frequency response due to the absorbed power ($P_{abs}$), and $\eta_{wb, bf}$ is the efficiency of wideband coupler before the filterbank.
Here, we used $\eta_{wb, bf} = -29$~dB, which was obtained by simulations using SONNET.
By dividing $x_{filter}$ by $x_{wb,bf}$, all systematic effects before the filterbank entrance were removed, and a pure filter transmission was extracted.
$x_{wb,bf}$ is the averaged response of the two wideband coupler KIDs before the filterbank.
In this way, the effect of a possible standing wave in the signal line could be cancelled as the two wideband couplers were placed in the distance of $\lambda_{eff}/4$ of the (sub-)mm wave signal in the signal line, resulting in a 90~degree phase shift and opposite sign for any standing wave.
The value of $\eta_{wb, bf} = -29$~dB, and the approximation of Eq.~\ref{Eq2}, which means the equality of the responsivity of all KIDs, were validated experimentally: the detail of how these values were obtained is discussed in Sec.~\ref{sensitivity}.

%Figure~\ref{Fig5} (b) is a zoom-in view of the response of one THz filter KID around the middle of the band ($\sim 300$~GHz).
%The red curve in the figure shows the result of a Lorentzian fit to the response, from which we extracted the resonance frequency ($F_{filter}$), the quality factor ($Q_{filter} = Q_i Q_c/(2Q_i+Q_c)$, $Q_c$: the coupling from the THz signal line (or KID) to the filter, $Q_i$: the internal quality factor of THz filter), and the maximum transmission ($|S_{31}|^2_{max}$) of the THz filter.
We applied a Lorentzian fit to each filter response, from which we extracted the filter frequency ($F_{filter}$), the quality factor ($Q_{filter} = Q_i Q_c/(2Q_i+Q_c)$, where $Q_c$ is the coupling from the THz signal line (or KID) to the filter, and $Q_i$ is the internal quality factor of THz filter), and the maximum transmission ($|S_{31}|^2_{max}$) of the THz filter.
Using $F_{filter}$s from the fitting, the filter responses were normalized in the frequency domain as well as the filter peak values, and the averaged (stacked) shape was calculated (Marting et al. in prep.), which is shown as the black curve in Fig.~\ref{Fig5} (b).
The colored curves represent the individual normalized filter responses.
The red dashed curve shows the result from a Lorentzian fitting to the averaged shape, and the obtained $Q_{filter}$ from the fit is $\sim 330$.
%We applied the fit to all filter responses, and Fig.~\ref{Fig5} (c) displays the obtained $Q_{filter}$ as a function of $F_{filter}$, showing an average of $340 \pm 50$, which is smaller than the design value of 500.
Figure~\ref{Fig5} (c) displays the obtained $Q_{filter}$ as a function of $F_{filter}$ from the fitting each individual filter responses, showing an average value of $340 \pm 50$.
Both results agree with each other, but show smaller $Q_{filter}$ than the design value of 500.
Assuming the same coupling $Q_c$ of the filter to the signal line and to the KID, we also estimated a coupling $Q_c \sim 1000$ and an internal $Q_i \sim 1200$ from the dips emerged in the transmission through the filterbank ($S_{21}$) calculated by $ |S_{21} (F)|^2 = x_{wb,af} (F) / x_{wb,bf} (F)$, where $x_{wb,af} (F)$ represents the averaged response of the two wideband coupler KIDs after the filterbank.
The detail of this analysis method is described in \citet{Laguna2021}.
We conclude that $Q_{filter}$ is lower than designed due to $Q_i$ being significantly lower than the value of $6000 - 10000$ from \citet{Buijtendorp2022}.
% discussion about frequency upshift??: L_s = 1.07 pH/sq --> 1.40 pH/sq, not sure why this occurred.

Figure~\ref{Fig5} (d)-(f) showcase the quality of the ISS chip.
The comparison of the 2-dimensional ($F_{filter},~f_r$) distribution between the measurement with the design is shown in Fig.~\ref{Fig5} (d).
%The ($F_{filter},~f_r$) distribution is designed based on the idea of: 1) separating $f_r$s of adjacent THz filters as far as possible to reduce KID-KID electromagnetic cross-talk; 2) but, at the same time, putting those filters relatively close in the $f_r$ space to reduce possible effect from fabrication such as film thickness and line width variations because these effects are spatially slow-varying quantities that can be minimized by placing KIDs $f_r$s not too far from each other \citep{Yates2014, Baselmans2017}. 
Both axes in the figure were normalized by the mean and standard deviation of $F_{filter}$ or $f_r$ to account for frequency shifts in the measured data.
We observe a good correspondence between the measurement and design values, while also highlighting a few missing ($F_{filter},~f_r$) pairs in the measurement dataset.
As a result, 334 out of 339 THz filters were identified, which means a yield of more than 98~\%.
Figure~\ref{Fig5} (e) displays spacing of the THz filter frequencies, $\Delta F_{filter} / F_{filter} = (F_{filter}^{i+1} - F_{filter}^{i}) / F_{filter}^i$, from the measurement.
We obtain a nearly constant value of $\Delta F_{filter}/F_{filter} = 0.02$ with a standard deviation of $2.7 \times 10^{-4}$, much lower than the filter spacing.
The residual slope on the line indicates a slow spatially varying systematic variation of a parameter determining the filter frequencies.
The frequency spacing of the filterbank can be expressed as a quality factor, $Q_{space}$, that is calculated as $Q_{space} = F_{filter} / \Delta F_{filter} = 500$, showing good agreement with the design value.
We also evaluated the frequency-scattering of KID, as shown in Fig.~\ref{Fig5} (f), by calculating the absolute deviation between measured and designed frequencies.
%We computed $|f_r^{measured} - f_r^{design}|/f_r^{design}$, yielding a $\sigma = 4.66 \times 10^{-4}$ by a Gaussian fitting, interestingly very similar to the filter frequency scatter.
We computed $|f_r^{measured} - f_r^{design}|/f_r^{design}$, yielding a $\sigma = 4.7 \times 10^{-4}$ by a Gaussian fitting, interestingly very similar to the filter frequency scatter.
This result was achieved through our two-step patterning process for the Al layer without employing trimming techniques \citep{Liu2017, Shu2018}, demonstrating the quality of the fabrication process \citep{Thoen2022}.

%The averaged $|S_{31}|^2_{max}$ from measurement was $0.161 \pm 0.038$, as indicated in the legend of Fig.~\ref{Fig5} (a).
The averaged $|S_{31}|^2_{max}$ from measurement was $0.16 \pm 0.04$, as indicated in the legend of Fig.~\ref{Fig5} (a).
This value is lower than the theoretical maximum coupling of a half-wavelength filter, which is 0.5.
Three factors contributed to this discrepancy.
The first cause was an internal oversampling.
The difference between $Q_{filter}=340$ and $Q_{space} = 500$ led to power being partially shared among neighboring filter channels.
This resulted in lower peak transmission of filters than the ideal case.
The second one was the variation in filter frequency $\Delta F_{filter} / F_{filter}$.
This further increased the effect of sharing power among neighboring channels.
The third effect is the low $Q_i$ of the filters, which results in power lost in the dielectric. 
To estimate the expected performance, we employed a numerical filterbank model \citep{Marting2024} using the measured oversampling and deviation values.
%This gave an expected value of $0.154 \pm 0.034$, which matched with the measured value.
This gave an expected value of $0.15 \pm 0.03$, which matches well with the measured value.

%                                  Two column figure
%__________________________________________________________________
 \begin{figure*}
   \centering
   \includegraphics[width=0.95\textwidth] {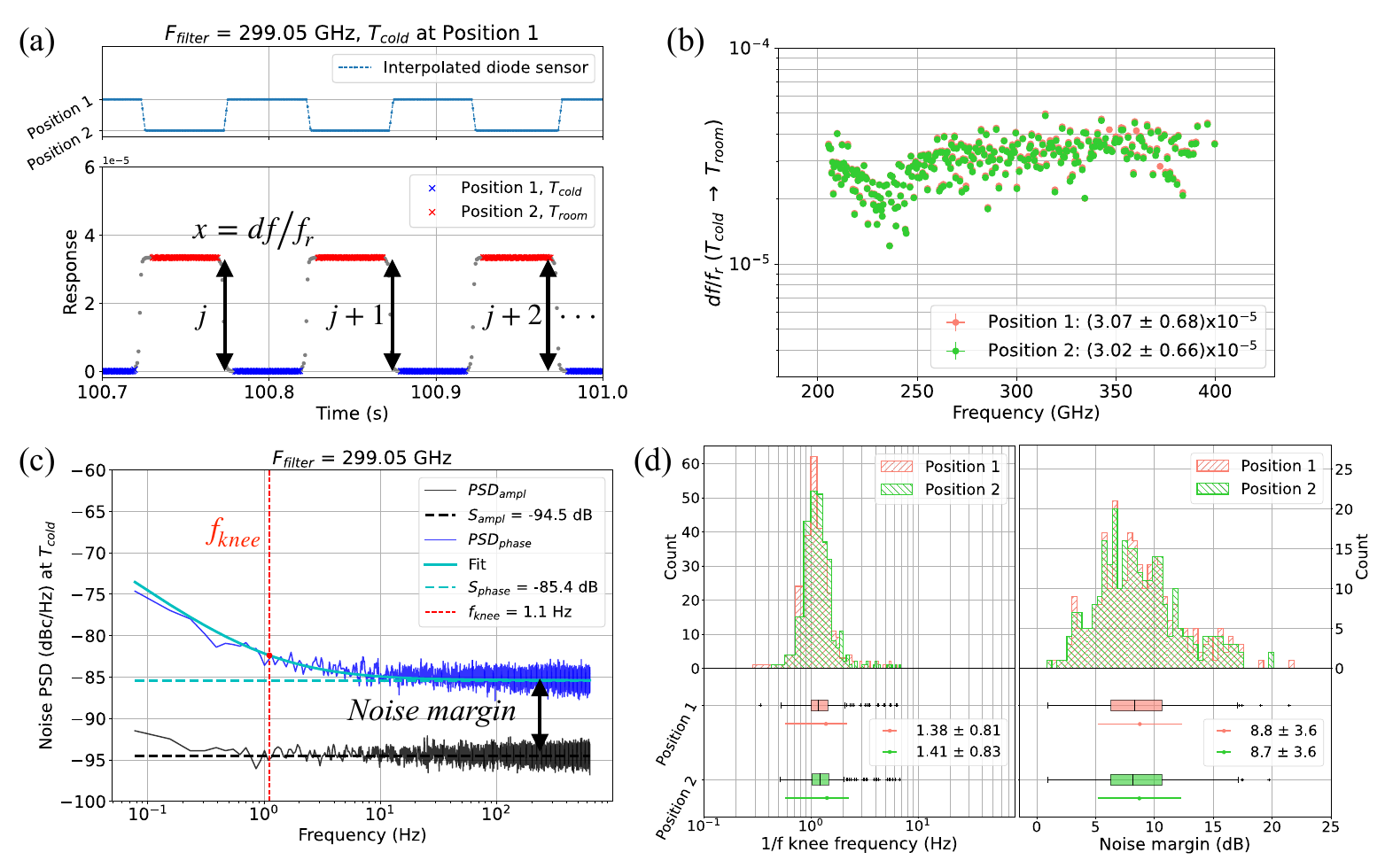}
   \caption{
   (a) An example of the time domain data of the $F_{filter} = 299.05$~GHz filter KID from the response measurement with the chopper, placing $T_{cold}$ at Position~1.
   The top panel shows the interpolated data from the diode sensor by the timestamp of the KID readout.
   The bottom panel displays the KID response: grey points are the KID response, while blue and red crosses represent the data points tagged as Position~1 ($T_{cold}$) and Position~2 ($T_{room}$) by the chopper diode data, respectively.
   (b) The obtained $x_{meas}$ as a function of $F_{filter}$ with averaged value and standard deviation in the legend.
   The pink and green points correspond to the data when placing $T_{cold}$ at Position~1 and Position~2, respectively.
   (c) An example of noise PSD of the $F_{filter} = 299.05$~GHz filter KID from the noise measurement.
   Blue and black curves represents the measured phase and amplitude noise PSD of the KID, respectively.
   The fit result to the phase PSD is shown as a cyan line, while dash lines represent the noise floor of the phase (cyan) and amplitude (black) PSDs.
   The vertical red dash line represents $f_{knee}$ of the phase noise PSD.
   (d) The left top panel displays the distribution of $f_{knee}$, with its box plot, averaged value, and standard deviation in the left bottom panel.
   The pink and green colors correspond to the data when placing $T_{cold}$ at Position~1 and Position~2, respectively.
   The right top panel displays the distribution of noise margin ($S_{phase} - S_{ampl}$ in dB unit), with its box plot, averaged value, and standard deviation in the right bottom panel.
   The color code is same as the left panel.
   }
              \label{Fig7}%
 \end{figure*}

%__________________________________________________________________
\subsection{Instrument sensitivity} \label{sensitivity}

Evaluation of instrument sensitivity involved two measurements: the KIDs response measurement using two black body radiators, one at room temperature ($T_{room} = 300$~K) and one at liquid nitrogen temperature ($T_{cold} = 77$~K), and noise measurement.
 During the response measurement, we placed two temperature loads at Position~1 and Position~2, indicated in Fig.~\ref{Fig4}, along the optical path of the chopper.
 We then measured the response of KIDs in the time domain by switching between these positions in 10~Hz through rotational motion of the chopper.
 In the noise measurement, 300~sec-long time domain data was taken with a sampling speed of $\sim 1.3$~kHz by fixing the chopper to the path where $T_{cold}$ was placed.
The noise power spectrum density (PSD) was then calculated from the time stream data to obtain the KID photon noise level and knee frequency of its $1/f$ noise.
From the measured KID response and noise PSD, we evaluated noise equivalent power (NEP) and optical efficiency of the whole instrument.

Figure~\ref{Fig7} (a) illustrates an example of how the KID response, $x = \delta f/f_r$, was extracted from the optical response measurement.
In this example, $T_{cold}$ and $T_{room}$ were placed at Position~1 and Position~2, respectively.
The top panel shows data from the diode sensor inside the chopper.
Note that we took the diode data with a sampling frequency around 800~Hz, which is lower than the KID readout speed of $\sim1.3$~kHz.
To address this mismatch, we interpolated the points by the timestamp of the KID readout, so that the diode data in the figure include intermediate states.
As an example, the bottom panel displays the filter KID response at $F_{filter} = 299$~GHz, showing good synchronization with the diode data.
The response for each KID was evaluated by taking the difference between two temperature loads.
With a sampling rate of $\sim1.3$~kHz and chopping frequency of 10 Hz, we observed approximately 130~points in one cycle.
However, due to transitions between the two chopper positions, about 24 points were discarded, which were determined using the diode data.
This resulted in less than 20~\% data loss overall, meaning an effective chopping efficiency of $\sim80$~\%.
To account for slow drifts (1/f noise), we calculated $x$ values at each cycle (denoted as $j,~j+1,~j+2...$ in the figure) and averaged them over the entire dataset: we write the averaged value as $x_{meas}$ hereafter.
Figure~\ref{Fig7} (b) shows the measured $x_{meas}$ as a function of $F_{filter}$.
The responses were almost identical when $T_{cold}$ was placed in either Position~1 or Position~2.
A small dip around $230 - 240$~GHz was due to the transmittance of the IR blocking filters stack shown in Fig.~\ref{Fig2} (c).

Figure~\ref{Fig7} (c) displays the measured noise PSD of the $F_{filter} = 299$~GHz filter KID.
The blue and black curves represent the phase and amplitude noise PSDs of the KID, respectively.
To characterize the noise properties of all KIDs, we fitted the noise PSDs with a function of the form: $a_0 + a_1/f$, where $a_0,~a_1$ are fitting parameters, and $f$ represents frequency in the readout domain.
This analysis allowed us to determine the knee frequencies ($f_{knee} \equiv a_1/a_0$) associated with 1/f noise and the white noise levels ($S_{ampl}~or~S_{phase} \equiv a_0$).
Note that the 1/f noise was mainly caused by TLS noise of the KIDs themselves so that we only evaluated $f_{knee}$ in the phase noise.
The distribution of $f_{knee}$ from all KIDs is plotted in the left panel of Fig.~\ref{Fig7} (d), showing an averaged $f_{knee}$ of about 1.4~Hz, which was well below the chopping frequency of 10 Hz, without any significant discrepancy between Position~1 and 2.
Furthermore, since $S_{ampl}$ was the same level as the readout electronics noise measured by the blind tones, higher values of $S_{phase}$ than $S_{ampl}$ indicated that the phase noise level was not limited by the readout noise: this was an important indication of the photon noise-limited sensitivity of the instrument.
To quantify this, we calculated a "noise margin" for each KID, defined as the difference between $S_{phase}$ and $S_{ampl}$ values in dB units  (i.e. $S_{phase} - S_{ampl}$ as indicated in Fig.~\ref{Fig7} (c)).
The distribution of noise margin is plotted in the right panel of Fig.~\ref{Fig7} (d).
The averaged noise margin across all KIDs was about 8.8~dB, which ensured the photon noise-limited sensitivity of the DESHIMA ISS system.
Note that a lower 1/f noise knee and higher photon noise level can be created by reducing the aluminium volume, this however goes at a cost to the KID dynamic range, which makes observations at varying sky opacities more cumbersome.

%                                  Two column figure
%__________________________________________________________________
   \begin{figure*}
   \centering
   \includegraphics[width=0.95\textwidth] {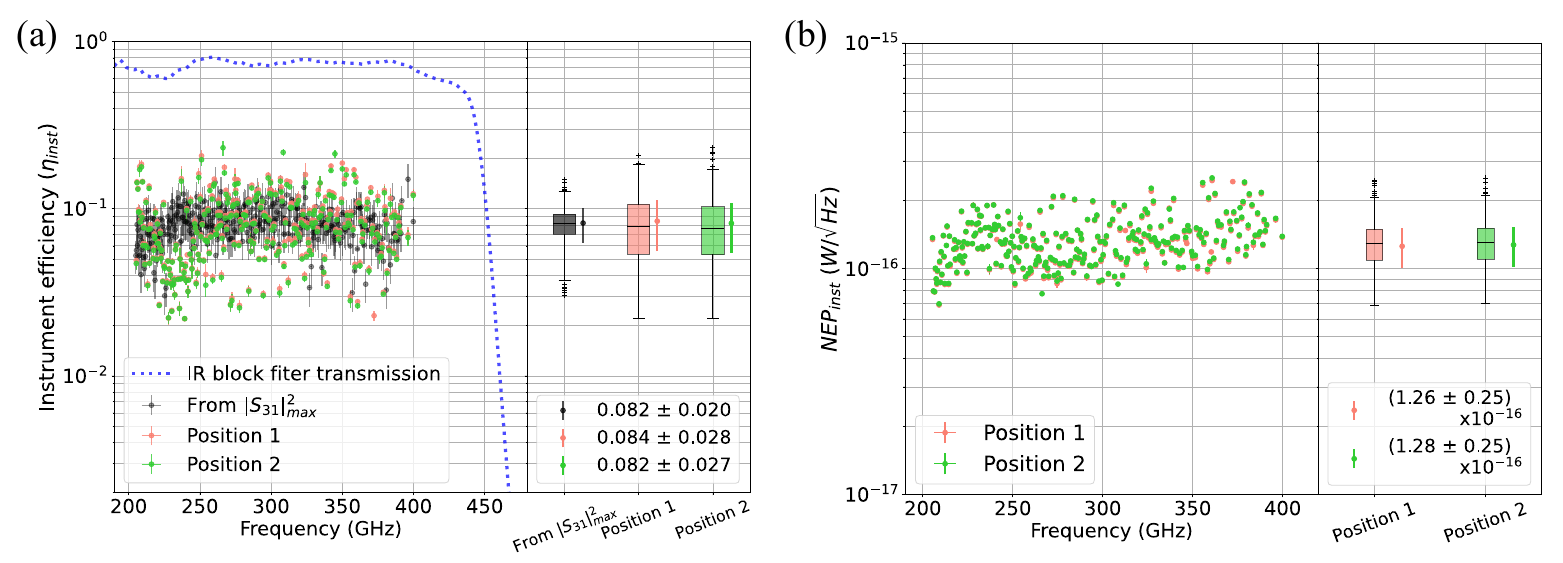}
   \caption{
   (a) The left panel shows instrument efficiency ($\eta_{inst}$) as a function of $F_{filter}$, with its box plot, averaged value, and standard deviation in the right panel.
   The pink and green colors correspond to the data when placing $T_{cold}$ at Position~1 and Position~2, respectively.
   The error bars of pink and green points were combination of statistical uncertainties from the response calculation and fitting error of the noise PSD.
   The black points are $\eta_{inst}$ calculated using $|S_{31}|^2_{max}$ in Fig.~\ref{Fig5} (a).
   The error bars were from uncertainties of $|S_{31}|^2_{max}$.
   The blue dot curve represents total transmission of IR blocking filters and Si window ($t_{IR} (F)$), same as shown in Fig.~\ref{Fig2} (c).
   (b) The left panel shows instrument optical NEP ($NEP_{inst}$) as a function of $F_{filter}$, with its box plot, averaged value, and standard deviation in the right panel.
   The pink and green colors correspond to the data when placing $T_{cold}$ at Position~1 and Position~2, respectively.
   }
   \label{Fig8}%
    \end{figure*}

Using the measured KIDs responses and the noise PSDs, NEP at loading temperature $T_{cold}$ can be determined experimentally by:
\begin{equation}
\centering
NEP_{exp}^i = \sqrt{S_x^i} \times \Bigg( \frac{dx^i}{dP_{rad}^i} \Bigg|_{T_{cold}} \Bigg)^{-1} \label{Eq3}.
\end{equation}
Here, $S_x$ represents the frequency noise, which can be derived from $S_{phase}$ using the relationship of $S_x = S_{phase} / 4 Q_r$, and $dx/dP_{rad}$ is the KID responsivity, representing how much the KID responses to changes in the loading power ($P_{rad}$).
The subscript $T_{cold}$ indicates that we use a readout tone at the KID resonant frequency at the lowest load temperature.

$P_{rad}$ could be defined at any reference plane.
In this paper, we define $P_{rad}$ as the loading power in front of the Si lens.
Then, $P_{rad}$ was calculated by:
\begin{eqnarray}
\centering
 P_{rad}^i (T_{load}) &=& \int dF~ \frac{1}{2} B(T_{load}, F) \lambda^2 T^i (F), \label{Eq4} \\
T^i (F) &=& t_{IR} (F) \times \Big( x_{filter}^{i,~norm} (F) ~/~ x_{wb, bf}^{norm} (F) \Big), \label{Eq5} %\\
%P_{abs}^i (T_{load}) &=& \eta_{opt}^i P_{rad}^i (T_{load}) \label{Eq5},
\end{eqnarray}
where $B(T, F)$ is the radiator irradiance, the factor $1/2$ shows the polarization efficiency, and $\lambda^2 =  (c/F)^2 = A \Omega$ is the total throughput, both representing a single polarization, single-mode detector.
$t_{IR} (F)$ is the total transmission of IR blocking filters shown in Fig.~\ref{Fig2} (c), and $x^{norm} (F) = x (F)~/ \max{ \{ x(F) \} }$ represents normalized filter response, where $x_{filter}(F)$ and $x_{wb,bf}(F)$ are taken from the THz frequency response presented in Sec.~\ref{freqresponse}.
In this way, we could extract the optical efficiency with respect to the Si lens, denoted as $\eta_{opt}$, by equating $NEP_{exp}$ to the theoretical photon-noise limited NEP as discussed below.
Hence  $\eta_{opt}$ represents combined efficiency of the leaky-wave antenna and a THz filter, and the absorbed power by the KID ($P_{abs}$) can be expressed as $P_{abs} = \eta_{opt} P_{rad}$.
For wideband KIDs, the factor $x^{i,~norm}_{filter}(F) / x^{norm}_{wb, bf}(F)$ in Eq.~\ref{Eq5} was replaced by the normalized frequency response from simulations using SONNET.
For dark KIDs, the factor was replaced by unity.

For calculation of the KID responsivity ($dx/dP_{rad}$), non-linear effects need to be considered especially when the KID response is measured using two very different temperature loads, such as $T_{cold}$ and $T_{room}$ \citep{Takekoshi2020}.
Since the KID response scales as $\sqrt{ P_{abs}} \propto \sqrt{P_{rad}} $ \citep{deVisser2014},  $x$ can be expressed as $x = k \sqrt{P_{rad}}$ with a proportional constant $k$.
Note that $x$ here is the KID frequency shift with respect to zero loading power, i.e. $x = (f - f_r^0)/f_r^0$ with $f_r^0$ representing the resonance frequency at zero loading power.
Therefore, in our measurement, $x_{meas} = (f_r^{T_{room}} - f_r^{T_{cold}}) / f_r^{T_{cold}}$ could be related with $k$ as:
\begin{equation}
x_{meas}^i \times f_r^{i, T_{cold}} / f_r^{i, 0} = k^i \Big( \sqrt{P_{rad}^i (T_{room})} - \sqrt{P_{rad}^i (T_{cold})}~\Big), \label{Eq6}
\end{equation}
so that the KID responsivity was calculated by:
%\begin{eqnarray}
\begin{align}
\frac{dx^i}{dP_{rad}^i} \Bigg|_{T_{load}} &=  \frac{k^i}{ 2 \sqrt{P_{rad}^i (T_{load})} } \notag \\
&= \frac{1}{ 2 \sqrt{P_{rad}^i (T_{load})} } \cdot \frac{ x_{meas}^i \times (f_r^{i, T_{cold}} / f_r^{i, 0}) }{ \sqrt{ P_{rad}^i (T_{room}) } - \sqrt{ P_{rad}^i (T_{cold}) } } \label{Eq7} \\
&\sim \frac{1}{ 2 \sqrt{P_{rad}^i (T_{load})} } \cdot \frac{ x_{meas}^i }{ \sqrt{ P_{rad}^i (T_{room}) } - \sqrt{ P_{rad}^i (T_{cold}) } }. \label{Eq7aprx}
%& \sim \frac{1}{ 2 \sqrt{P_{rad}^i (T_{load})} } \cdot \frac{ x_{meas}^i }{ \sqrt{ P_{rad}^i (T_{room}) } - \sqrt{ P_{rad}^i (T_{cold}) } }.
	%k^i &=&  \frac{ x_{meas}^i \times (f_r^{i, T_{cold}} / f_r^{i, 0}) }{ \sqrt{ P_{rad}^i (T_{room}) } - \sqrt{ P_{rad}^i (T_{cold}) } }. \label{Eq7}
%\end{eqnarray}
\end{align}
%Although measuring $f_r^0$ was quite impossible, we could approximate $f_r^0$ by the KID resonance frequency measured in the dark condition by closing the optical window (denoted as $f_r^{dark}$).
We can approximate $f_r^0$ by the KID resonance frequency measured in the dark condition by closing the optical window (denoted as $f_r^{dark}$).
The difference between $f_r^{dark}$ and $f_r^{T_{cold}}$ was measured to be  less than 1~MHz, which justified the approximation of $f_r^{T_{cold}} / f_r^0 \sim 1$ within 0.1~\% uncertainty because KID resonance frequencies were in the range of $4 - 6$~GHz.

The optical efficiency with respect to the Si lens, $\eta_{opt}$, is then calculated with the assumption of the photon noise limited sensitivity.
By equating the calculated $NEP_{exp}$ to the theoretical optical NEP of photon-noise limited KIDs \citep{Ferrari2018}, we obtained $\eta_{opt}$ as:
\begin{eqnarray}
%\eta_{opt}^i &=& \frac{ 2 \int dF~\frac{1}{2} B(T_{load}, F) \lambda^2 T^i (F) \times hF + 4 P_{rad}^i \Delta / \eta_{pb} }{ (NEP_{exp}^i)^2 - 2 \int dF~\Big( \frac{1}{2}B(T_{load}, F) \lambda^2 \Big)^2 T^i (F) } \label{Eq6} \\
%&\sim& \frac{ 2 P_{rad}^i hF_{filter}^i + 4 P_{rad}^i \Delta / \eta_{pb} }{ (NEP_{exp}^i)^2 - 2 (P_{rad}^i)^2 / \Delta F^
%\label{Eq7}
\eta_{opt}^i &=& \frac{ 2 P_{rad}^i hF_{filter}^i + 4 P_{rad}^i \Delta / \eta_{pb} }{ (NEP_{exp}^i)^2 - 2 (P_{rad}^i)^2 / \Delta F^i}, \label{Eq8}
\end{eqnarray}
where $h$ is the Planck constant, $\Delta$ is the superconducting gap energy of Al film, $\eta_{pb} \sim 0.4$ is the pair-breaking efficiency \citep{Guruswamy2014}, and $\Delta F^i$ represents equivalent bandwidth calculated by $\Delta F^i = \int dF~T^i (F)$.
As mentioned above, $\eta_{opt}$ is a product of a THz filter efficiency, $\eta_{filter}$, and the leaky-wave antenna efficiency, $\eta_{leaky}$, i.e. $\eta_{opt}^i = \eta_{filter}^i \times \eta_{leaky}$, where $\eta_{leaky}$ further breaks down into spill-over from the antenna beam on the lens surface, fraction of the power radiated to the front side of the lens, fraction of the power coupled to the CPW signal line, reflection at the lens surface, and spill-over from the cold mirrors.
In total, $\eta_{leaky}$ is $\sim 0.7$ by the design (Dabironezare et al. in prep.).

The instrument efficiency ($\eta_{inst}$) and NEP ($NEP_{inst}$), which we defined in this paper as the efficiency and NEP with respect to the entrance of the instrument (MP~2 in Fig.~\ref{Fig4}, i.e. the entrance of chopper), were calculated by the following equation:
\begin{eqnarray}
\centering
\eta_{inst}^i &=& \eta_{opt}^i \times t_{IR} (F_{filter}^i) \notag \\
&=& \eta_{filter}^i \times \eta_{leaky} \times t_{IR} (F_{filter}^i), \label{Eq9} \\
NEP_{inst}^i &=& NEP_{exp} ^i \times \Big( t_{IR} (F_{filter}^i) \Big)^{-1}. \label{Eq10}
\end{eqnarray}
Here, we assumed that the loss in the warm optics chain, such as spill-over loss from the warm mirrors or the chopper, was negligible.
In order to check this assumption, we also did the optical sensitivity measurement at MP~1 (in front of the cryostat window, see Fig.~\ref{Fig4}), but without chopper, by exchanging $T_{room}$ and $T_{cold}$ alternately in a time scale of a few 10s~second.
The obtained $\eta_{opt}$ values were consistent. %with each other within a few~\% level.

Figure~\ref{Fig8} (a), (b) display obtained $\eta_{inst}$ and $NEP_{inst}$ as a function of $F_{filter}$, respectively.
Both quantities show quite uniform distribution over the frequency band.
The averaged $\eta_{inst}$ was $\sim 8$~\% while the averaged $NEP_{inst}$ was $\sim 1.3 \times 10^{-16}$~W/$\sqrt{Hz}$, without any significant difference between Position~1 and Position~2.
In Fig.~\ref{Fig8} (a), we also plotted $\eta_{inst}$ calculated using $|S_{31}|^2_{max}$ in Fig.~\ref{Fig5} (a), which was an equivalent quantity to $\eta_{filter}$, by black points as reference.
The consistency between two separate results (from Step~1 and Step~4 in Sec.~\ref{labsetup}) increase the reliability of the obtained results.

\begin{figure}
   \centering
    \includegraphics[width=0.5\textwidth] {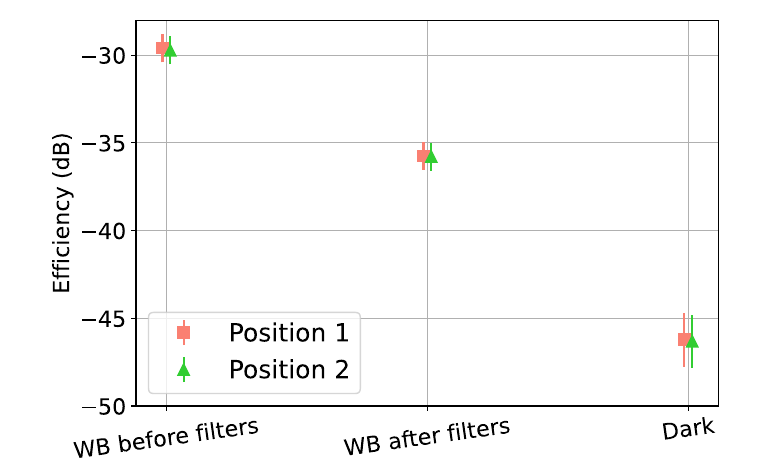}
      \caption{
      The measured efficiencies of wideband couplers before (left row) and after (middle row) the filterbank, and dark KID (right row).
      The number of KIDs of wideband couplers before, after, and dark are 2, 2, and 4, respectively, but averaged values are plotted here.
      The error bars were combination of variation when averaging in each group, statistical uncertainties from the response calculation, and fitting error of the noise PSD.      
      The pink squares and green triangles correspond to the data when placing $T_{cold}$ at Position~1 and Position~2, respectively.
              }
         \label{Fig9}
   \end{figure}

Along with the instrument efficiency, we also evaluated efficiency of wideband coupler KIDs and dark KIDs as shown in Fig.~\ref{Fig9}.
In the plot, contribution from $\eta_{leaky}$ was removed by calculating the ratio $\eta_{opt}/\eta_{leaky}$ to evaluate purely wideband coupler coupling and stray light effects.
%As a result, the averaged coupling strength of the wideband coupler KIDs before ($\eta_{wb, bf}$) and after ($\eta_{wb, af}$) the filterbank were measured to be $-29.6 \pm 0.8$~dB and $-35.7 \pm 0.8$~dB, respectively.
As a result, the averaged coupling strength of the wideband coupler KIDs before ($\eta_{wb, bf}$) and after ($\eta_{wb, af}$) the filterbank were measured to be $-30 \pm 1$~dB and $-36 \pm 1$~dB, respectively.
The results showed that $\eta_{wb, bf}$ was consistent with the design of $-29 \pm 1$~dB, while $\eta_{wb, af}$ was $6-7$~dB lower, indicating that the power was reduced by a factor of 4 to 5 due to power absorbed in the filterbank.
%The averaged coupling strength of the dark KIDs was $-46.2 \pm 1.5$~dB, which was about $-38$~dB lower than $\eta_{filter}$ (average $\sim 0.16 = - 8$~dB), showing that the stray light effect on the KID detectors was very small.
The averaged coupling strength of the dark KIDs was $-46 \pm 2$~dB, which was about $-38$~dB lower than $\eta_{filter}$ (average $\sim 0.16 = - 8$~dB), showing that the stray light effect on the KID detectors was very small.

%                                  One column figure
%__________________________________________________________________

%%__________________________________________________________________
%\subsection{Beam characteristics} \label{beampattern} 

%                                  Two column figure
%__________________________________________________________________
   \begin{figure*} 
   \centering
   \includegraphics[width=0.9\textwidth] {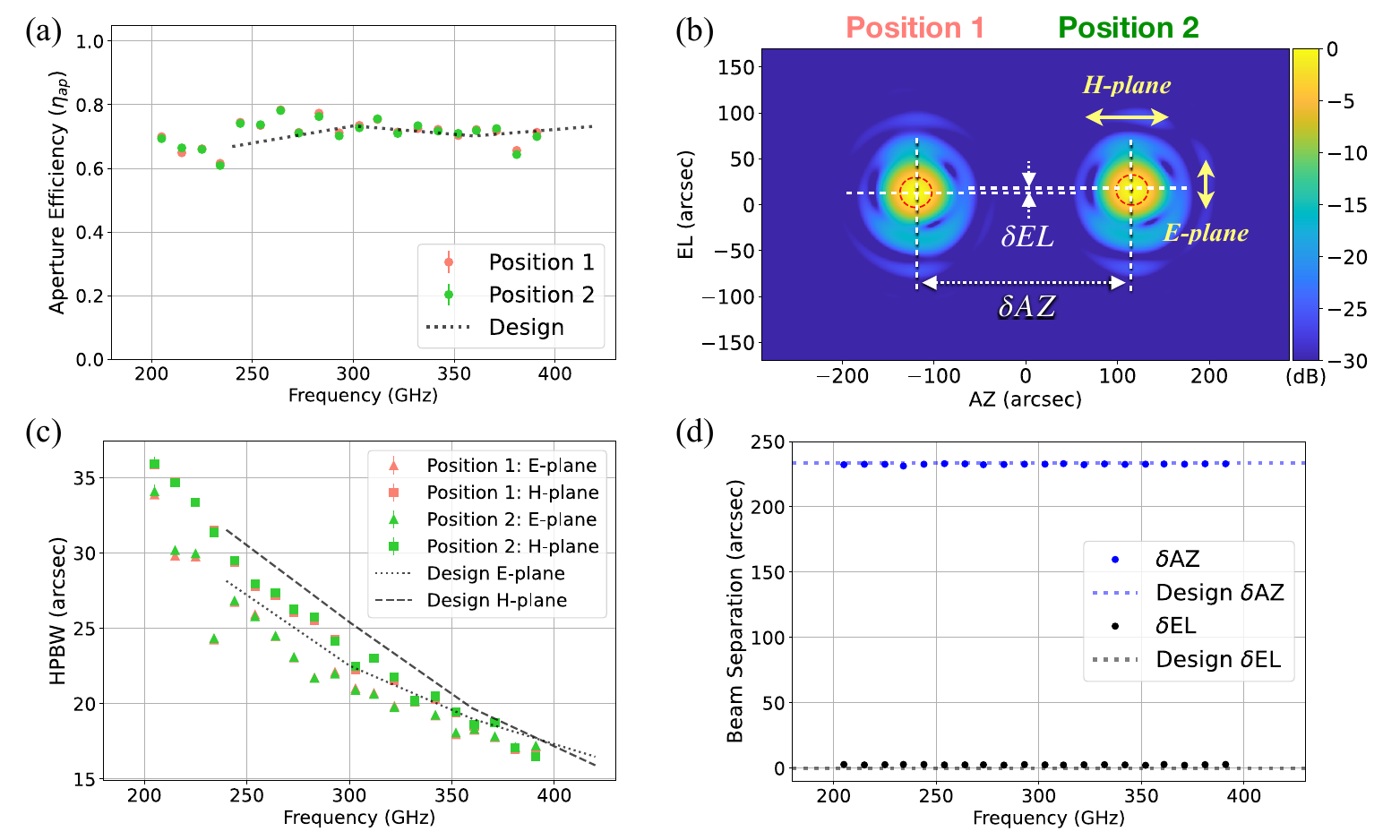}
      \caption{
      The expected performance of the DESHIMA~2.0 instrument on the ASTE telescope based upon the laboratory PA beam pattern measurements.
      (a) The obtained aperture efficiency ($\eta_{ap}$) as a function of filter frequency (20 filters matched to the harmonic frequencies).
      The pink and green colors correspond to the beam pattern data measured at Position~1 and Position~2, respectively.
      The black dot curve represents the design specification.
      (b) The telescope far-field beam patterns of the 205~GHz filter at both paths of Position~1 and 2. %, calculated by propagating the measured beam patterns through a telescope model using PyPO.
      The red dash lines plot 3~dB contours from the results of 2-dimensional Gaussian fittings.
      The definitions of E and H planes are indicated by yellow arrows: the E-plane is aligned along the semi-minor axis of the far-field main beam, while the H-plane is parallel to the semi-major axis.
      The definitions of beam separation ($\delta AZ,~\delta EL$) are also illustrated in the figure.
      (c) The obtained HPBW in the E-plane (triangle points) and H-plane (square points) as a function of filter frequency.
      The pink and green colors correspond to the beam pattern data measured at Position~1 and Position~2, respectively.
      The black dot and dash curves represent the design specification of E-plane and H-plane, respectively.
      (d) The beam separation as a function of filter frequency.
      The blue and black points are evaluated $\delta AZ$ and $\delta EL$, respectively, with the design value plotted as dot lines.      
              }
         \label{Fig10}
   \end{figure*}

%__________________________________________________________________
\subsection{Beam characteristics} \label{beampattern}

We measured both the phase and amplitude (PA) beam patterns of the instrument at MP~2 in Fig~\ref{Fig4}, following a quasi-heterodyne measurement technique  \citep{Davis2019, Yates2020}, which allowed us to obtain PA beam patterns using any KID in the ISS as detector.
The PA information was essential to investigate misalignment of the instrument, and also propagate the measured beam pattern numerically through the telescope reflectors to estimate expected performance of the instrument at the telescope.
The experimental procedure is similar to \citet{Endo2019a}, but we extended the PA measurement to a so-called the harmonic phase and amplitude (HPA) measurement.
 This uses two broadband mixers \citep{Paveliev2012} to generate overtones at integer multiples of the pump signal of $f_{synth} \sim 9.76548$~GHz, which were fed from two synthesizers connected to the mixers.
 The frequency of generated radiation was $n \times f_{synth}$, and the integer range of $n = 21 - 40$ corresponded to the frequency range of the DESHIMA~2.0.
 The two synthesizers had a small frequency difference of $\Delta f \sim 12$~Hz, resulting in a THz frequency offset of $\Delta F = n \times \Delta f$ that modulated the detector response in the time domain.
 $\Delta f$ was chosen in such a way that the $\Delta F$ with $n=40$ was well within the bandwidth of the detectors and readout system with a sampling speed of $\sim1.3$~kHz.
 In this way, the KIDs responses at the modulation frequency of $\Delta F$ in the readout domain could be correlated to the incoming source signal of the frequency of $n \times f_{synth}$.
 As KIDs are not phase-sensitive detectors, the modulation of $\Delta f$ was separately measured by adding it to all the readout tones in the multiplexing readout: the phase of this modulation could be used as a phase reference.
 The complex field parameters (phase and amplitude of the source signal) were then obtained by the values at the frequency of $\Delta F$s of a complex fast Fourier transform (FFT) of the time-domain data by normalising the FFT phase by the phase reference. 
 While one mixer is fixed, the other mixer can be moved in a scanning plane, allowing to map PA beam patterns as a function of the scanning position across the harmonic range of THz frequencies simultaneously.
 More details of the experiment are given in \citet{Moerman2024}.
 As the origin of the harmonic frequencies are a well calibrated signal from the synthesizers, we also used this data for the absolute frequency calibration of the THz filter frequencies, which is discussed in Sec.~\ref{freqcalibration}.

We used the measured PA beam patterns to estimate the expected performance of the instrument at the telescope by propagating them through a model of ASTE using PyPO. %, which also supported physical optics calculation.
Note that this study was limited to 20 filters, whose filter frequencies matched the harmonic frequencies in the HPA measurement, i.e. $n \times f_{synth}$ with integer values ranging from $n = 21$ to $40$.
These results are shown in Fig~\ref{Fig10}.

We first evaluated the antenna efficiencies, consisting of the spillover efficiency on the sub-reflector ($\eta_{so}^{sub}$), taper efficiency in the aperture of the main-reflector  ($\eta_t$), and aperture efficiency $\eta_{ap} = \eta_{so}^{sub} \times \eta_t$. 
Figure~\ref{Fig10} (a) displays the obtained $\eta_{ap}$ for both beam positions after the chopper.
We find an average value of $\sim 0.7$, which agrees well with the design value indicated by the dashed line in the figure, while also showing no significant difference between Position~1 and 2.
We also obtained $\eta_{so}^{sub} \sim 0.9$, which is a bit lower than the designed value of 0.95, and $\eta_t \sim 0.8$, which is a bit larger than the design value of 0.75.
The combined effect is a slight increase in the edge illumination on the sub-reflector compared to the design.
This also results in a higher illumination at the outer section of the main-reflector.
%By design, the main-reflector was intentionally under-illuminated for the DESHIMA 2.0 instrument.
Consequently, while the increased illumination at the edge does not significantly impact the instrument performance (as evidenced by good agreement between measured and designed $\eta_{ap}$ values), it results in a decrease in beam size in far-field, which was consistent with smaller half-power beam widths (HPBWs) than originally specified as displayed in Fig.~\ref{Fig10} (c).
% ### mention about z focus??:
% ### As our alignment strategy does not explicitly correct for defocus, which is misalignment along the optical axis, the sub-reflector is used for refocusing the optical system after the warm optics alignment at the telescope. 
Here, $\eta_t$ was calculated in the aperture plane perpendicular to the telescope pointing, which was in the direction of maximum directivity and off from the broadside in the azimuthal direction due to the two separated paths (Position~1 and 2) in the sky chopper.
For more details of the definition and calculation of those efficiencies, see Appendix of \citet{Moerman2024}.
We also note that $\eta_{ap}$ here does not contain the surface error of the ASTE telescope.
The optical verification of the DESHIMA~2.0 instrument at ASTE, including the surface error, has been done and details are described in \citet{Moerman2025}.

An example of the expected telescope far-field beam pattern of the 205~GHz filter (corresponding to the 21-st harmonic frequency) is displayed in Fig.~\ref{Fig10} (b).
%We propagated the measured beam patterns in both Position~1 and 2 separately using PyPO.
As the coordinates of the measurement plane (MP~2) was properly implemented in the telescope model through the alignment process, the far-field beam patterns from Position~1 and 2 showed an offset from the telescope broadside ($AZ=EL=0$) as expected.
To extract the beam characteristics such as pointing (beam center) and half-power beam widths (HPBW), we fitted the far-field beam patterns to a 2-dimensional Gaussian function.
Figure~\ref{Fig10} (c) displays the obtained HPBWs in both the E-plane and H-plane as a function of filter frequency, again showing no significant discrepancy between Position 1 and 2. 
The definition of the E and H planes are illustrated in Fig.~\ref{Fig10} (b).
The HPBWs were slightly smaller than the design values across all measured frequencies.
As discussed above, we attributed this discrepancy to a higher edge illumination on the sub-reflector, which could also explain the lower $\eta_{so}^{sub}$ and larger $\eta_t$.
Furthermore, using the obtained beam centers from the fitting, we also estimated the beam separation between the far-field beam patterns of Position~1 and 2.
We calculated $\delta AZ$ and $\delta EL$, which definition are also illustrated in Fig.~\ref{Fig10} (b), and plotted in Fig.~\ref{Fig10} (d) as a function of filter frequency.
The averaged $\delta AZ$ and $\delta EL$ were $232.6 \pm 1.5$ and $2.6 \pm 1.9$~arcsec, respectively, which were consistent with the design specification of $\delta AZ = 234$ and $\delta EL = 0$~arcsec.

%%__________________________________________________________________
%\subsection{Absolute frequency calibration} \label{freqcalibration} 

%                                  One column figure
%__________________________________________________________________
   \begin{figure*}
   \centering
   \includegraphics[width=0.9\textwidth] {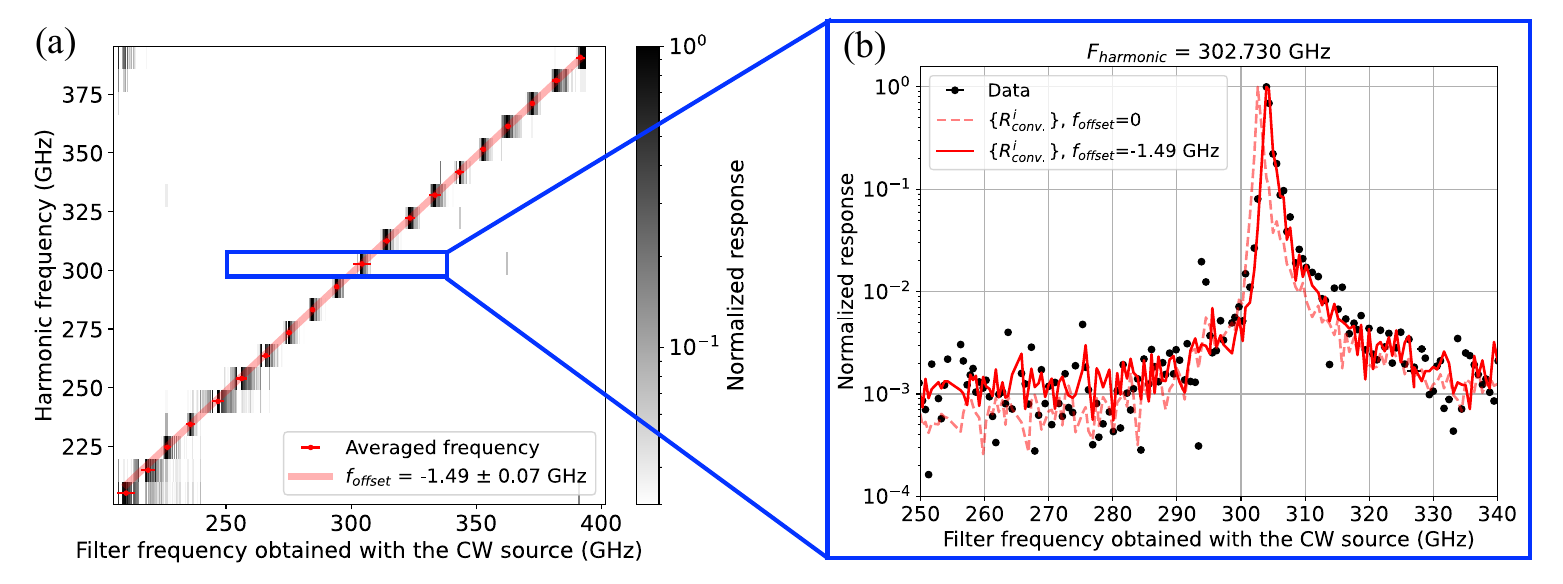}
   \caption{
   	(a) 2-dimensional plot of the normalized KIDs responses obtained from the HPA beam pattern measurement.
	The vertical axis shows harmonic frequency, while the horizontal axis presents filter frequency measured with the CW source.
	Due to the second resonances of the lower frequency filters (filter peaks around $200-220$~GHz), they also respond to signals from high harmonic frequencies (black points around the top left of the figure).
	The red points represent the weighted average of the filter frequencies by the KIDs responses, and the red curve shows the result from a linear fit.
	The blue box in the figure indicates the plotted data in the panel (b).
	(b) Comparison of the measured data (black points) and the convolved response of $\{ R_{conv.}^i \}~/ \max \{ R_{conv.}^i \}$ for the harmonic frequency of 302.730~GHz.
	The red dash and solid curve are convolved responses with $f_{offset}=0$ and $-1.49$~GHz, respectively.
	}
              \label{Fig12}%
    \end{figure*}
%
%

%__________________________________________________________________
\subsection{Absolute frequency calibration} \label{freqcalibration} 

Because of a frequency uncertainty of 2~GHz from the CW source used in the Step~1 experiment, we carried out an absolute frequency calibration of the filters using the data from the Step~3 experiment in which the harmonic frequencies were generated by well-calibrated signals from a synthesizer.

Figure~\ref{Fig12} (a) illustrates this method.
The vertical axis shows the frequency generated by means of the synthesizer and harmonic mixers, while the horizontal axis presents filter frequency measured with the CW source.
The grey scale points in the figure show the KIDs responses normalized to the maximum response at each frequency generated by the harmonic mixer, and the red points represent the weighted average of the filter frequencies by the KIDs responses.
We fitted the red points by a linear function with the intercept as a fitting parameter (denoted as $f_{offset}$), while fixing the slope to unity.
As a result, we obtained $f_{offset} = -1.49 \pm 0.07$~GHz, which was consistent with the 2~GHz accuracy of the CW source, demonstrating an improvement of the absolute frequency calibration of the filterbank with the data from the HPA beam pattern measurement.
Allowing also for the slope of the line as fit parameter did not change the result, indicating we can use $f_{offset}$ as only parameter to correct the DESHIMA~2.0 absolute filter frequencies.

Figure~\ref{Fig12} (b) shows how the obtained $f_{offset}$ improved the matching between the measured data (black points, normalized to its maximum) and the convolved responses (red curves).
%Given an $f_{offset}$ value, expected response of the DESHIMA instrument could be calculated by convolving a harmonic frequency with the filter responses by the following formula:
The convolved responses were calculated by the following formula:
\begin{eqnarray}
\centering
R_{conv.}^i &=& \int dF~T^i(F + f_{offset})~\delta(F - F_{harmonic}) \label{Eq11} \\
 &=& T^i (F_{harmonic} + f_{offset}),  \label{Eq12}
\end{eqnarray}
where i denotes $i$-th filter, $T^i (F)$ is the transmission of filter $i$ (same as in Eq.~\ref{Eq5}), $\delta (F)$ represents the Dirac delta function, and $F_{harmonic}$ is the harmonic frequency from the HPA beam pattern measurement.
We used the delta function in the convolution because the tone bandwidth from the harmonic mixer was much narrower than $Q_{filter}$, resulting in $R_{conv.}^i$ equal to the interpolated $i$-th filter transmission value at $F = F_{harmonic} + f_{offset}$.
%Figure~\ref{Fig12} (b) shows how the obtained $f_{offset}$ improved the matching between the measured data (normalized to its maximum) and the convolved response of $\{ R_{Conv.}^i \}~/ \max \{ R_{Conv.}^i \}$ for the harmonic frequency of 302.730~GHz (corresponding to the 31-st harmonic frequency).
The red curves in the figure shows two sets of $\{ R_{conv.}^i \}$ of all $i$-th filters at $F_{harmonic} = 302.730$~GHz with $f_{offset} = -1.49$~GHz (red solid curve) and $f_{offset} = 0$ (red dash curve), respectively (both normalized by the maximum value in each set).
The convolved response with $f_{offset} = -1.49$~GHz shows better agreement with the measured data (black points), especially around the peak.
%### Affect to Q value:
%### Q = F0/dF --> dQ/Q = sqrt( (dF0/F0)**2 + (ddF/dF)**2 ) = abs(dF0/F0)
%### dF0 = -1.5 GHz, F0 = 200-400 GHz --> dQ/Q = 0.008 - 0.004 < 1 %
%### The relative frequency accuracy of 20~MHz is ignored as it is less than the error of $Q_{filter}$: need to be quantified??

%______________________________________________________________
\section{Conclusions} \label{conclusion}

We presented extensive characterizations of the DESHIMA~2.0 instrument in the laboratory measurements.
The measured frequency coverage was $200 - 400$~GHz with a mean $Q_{filter}$ of $340 \pm 50$, lower than the design value of $Q_{filter} = 500$, which is caused by a low MS filter $Q_i \sim 1200$, and instrument efficiency of $\sim 8$~\%, indicating 4 times wider band coverage and 4 times higher coupling efficiency than DESHIMA~1.0.
The filter separation was found to be $Q_{space} = F_{filter} / \Delta F_{filter} = 500$, exactly as designed, with a very low scatter given by $\sigma_{\Delta F_{filter} / F_{filter}} = 2.7 \times 10^{-4}$, showing excellent control of the fabrication process.
We identified 334 out of 339 THz filters, which means an yield rate of more than 98~\%.
The KID phase noise showed the photon noise limited sensitivity with a mean $f_{knee}$  of 1.4~Hz, which is well below 10~Hz of the chopping speed of the sky-position chopper.
The data loss caused by the chopper was less than 20~\%, and the telescope far-field beam patterns through the chopper matched the calculated field distributions exactly, importantly showing no truncation by the chopper construction.
Also the beam separation for the two chopper positions matched the design well.
The estimated aperture efficiency was found to be in good agreement with the designed value of approximately 70~\%.
We also demonstrated validity of a new method of absolute frequency calibration using the data from the HPA beam pattern measurement.

%On the other hand, there was a discrepancy between $Q_{filter}$ and $Q_{space} = 500$ that caused an internal oversampling, and hence led to a lower filter efficiency than the case of the ideal frequency sampling (i.e. $Q_{filter} = Q_{space}$).
%The lower $Q_{filter}$ than the design value of 500 was due to the fact that $Q_i$ was lower than the expectation.
To improve instrument efficiency in future developments, a few potential strategies could be explored:
i) Increasing the $Q_i$ of the dielectric by using a different material \citep{Defrance2025},
ii) matching $Q_{filter}$ and $Q_{space}$ from the current $Q_{space} > Q_{filter}$, which causes signal leakage to neighboring filters, and
iii) using directional filters, which offer significantly higher in-band transmission (theoretical maximum in-band coupling of unity \citep{Marting2024}, compared to $\sim0.5$ for our current filter design).
%Additionally, other promising avenue with the DESHIMA ISS technology is to develop an integrated field unit (IFU) by increasing the number of spatial pixels.
%\textcolor{red}{
These developments will bring up the instrument efficiency comparable with heterodyne receivers.
Together with the scalability to IFUs with  $\gtrsim 100$ spatial pixels, the ISS technology will lead to unique astronomical instruments complementary to the existing large scale interferometers.
%}

%__________________________________________________________________
\begin{acknowledgements}
      This work was supported by the European Union (ERC Consolidator Grant Nos. 648135 MOSAIC and 101043486 TIFUUN).
      Views and opinions expressed are however those of the authors only and do not necessarily reflect those of the European Union or the European Research Council Executive Agency.
      Neither the European Union nor the granting authority can be held responsible for them.
      This research was also supported by the Netherlands Organization for Scientific Research NWO (Vidi Grant No. 639.042.423, NWO Medium Investment Grant No. 614.061.611 DESHIMA, and Open XS grant OCENW.XS22.4.146),
      A.P.L. was supported by the Juan de la Cierva grant JDC2023-051842-I funded by the Spanish MCIN/AEI/10.13039/501100011033.
      M.R. is supported by the NWO Veni project "Under the lens" (VI.Veni.202.225).
      T.T. was supported by the MEXT Leading Initiative for Excellent Young Researchers (Grant No. JPMXS0320200188).
      This study was also supported by the JSPS KAKENHI (Grant Nos. JP22H04939, JP23K20035, JP24H00004, and JP24H01808).
      %The ASTE telescope is operated by the National Astronomical Observatory of Japan (NAOJ).
\end{acknowledgements}

\bibliographystyle{aa} % style aa.bst
\bibliography{reference} % your references Yourfile.bib

\end{document}